\documentclass[apj]{emulateapj}
\usepackage{graphicx}
\usepackage{amsmath}
\usepackage{multirow}
\usepackage{subfigure}
\begin{document}

\slugcomment{Accepted to ApJ December 20, 2011}

\title{Warm Absorbers and Outflows in the Seyfert-1 Galaxy NGC 4051}
\author{Ashley L. King\altaffilmark{1}, Jon M. Miller\altaffilmark{1}, John Raymond\altaffilmark{2}}

\altaffiltext{1}{Department of Astronomy, University of Michigan, 500 Church Street, Ann Arbor, MI 48109, ashking@umich.edu}
\altaffiltext{2}{Harvard-Smithsonian Center for Astrophysics, 60 Garden Street, Cambridge, MA 02138, USA}

\journalinfo{The Astrophysical Journal}

\begin{abstract}
We present both phenomenological and more physical photoionization models of the Chandra HETG spectra of the Seyfert-1 AGN NGC 4051. We detect 40 absorption and emission lines, encompassing highly ionized charge states from O, Ne, Mg, Si, S and the Fe L-shell and K-shell. Two independent photoionization packages, XSTAR and Cloudy, were both used to self-consistently model the continuum and line spectra. These fits detected three absorbing regions in this system with densities ranging from 10$^{10}$ to 10$^{11}$ cm$^{-3}$. In particular, our XSTAR models require three components that have ionization parameters of $\log \xi =$ 4.5, 3.3, \& 1.0, and are located within the BLR at 70, 300, and 13,000 R$_g$, respectively, assuming a constant wind density. Larger radii are inferred for density profiles which decline with radius. The Cloudy models give a similar set of parameters with ionization parameters of $\log \xi$ = 5.0, 3.6, \& 2.2 located at 40, 200, and 3,300 R$_g$.  We demonstrate that these regions are out-flowing from the system, and carry a small fraction of material out of the system relative to the implied mass accretion rate. The data suggest that magnetic fields may be an important driving mechanism.

\end{abstract}

\section{Introduction}
The advent of CCD spectroscopy with ASCA showed that the ``warm absorbers" \citep{Halpern84} that are present in the X-ray spectra of some Seyfert galaxies can be described with a combination of He-like O VII and H-like O VIII absorption edges \citep{Reynolds97}.  The spectral resolution of CCDs is insufficient to clearly reveal individual narrow absorption lines that should be associated with such edges.  The high resolution dispersive spectrometers aboard {\it Chandra} and {\it XMM-Newton} have been able to clearly reveal such narrow absorption lines in a number of Seyfert AGN.  Indeed, it is now clear that X-ray warm absorbers span a range of ionizations, generally have blue-shifts on the order of $10^{2-3}~{\rm km}~{\rm s}^{-1}$ with respect to their host galaxy, and so must be outflows \citep{Kaastra00, Kaspi00, Kaspi02}.

With deep observations and careful, self-consistent photoionization studies of high resolution spectra, the nature and constraints on the origin of these outflows can -- in principle -- be determined.  In some warm absorbers, there is evidence that the opacity must derive partly from elements trapped on dust grains \citep{Lee01}, suggesting that the absorption may partially originate in a parsec--scale ``torus" \citep{Antonucci93}.  Similarly distant absorption is a viable description of the absorbers in many spectra \citep[see, e.g.][]{ Blustin05}.  These outflows are plausibly powered by radiation pressure from the central engine.  In this picture, the outflows may be termed ``disk winds", but they are unlikely to provide deep insights about the nature of disk accretion for they lack the ability to transfer angular momentum out of the system.

The best spectra of Seyfert warm absorbers may suggest a different origin.  The 900~ksec {\it Chandra}/HETG spectrum of NGC 3783 is richly detailed, and photoionization analysis suggests that it may originate within several hundred gravitational radii (${\rm r}_{g} = {\rm GM}/{\rm c}^{2}$) of the black hole \citep{Kaspi02}. Detailed spectral modeling of the X-ray absorption in NGC 4151 finds high column densities and evidence of large bulk motion that may be require large non-radial motions \citep{Kraemer05}.  This is qualitatively consistent with magnetocentrifugal disk winds of the kind described by \cite{Blandford82}.  Owing to the fact that these winds can remove angular momentum from accreting gas (thus allowing it to accrete), some warm absorbers (or the most ionized, innermost regions of warm absorbers) may hold clues to the fundamental physics of disk accretion \citep[see, e.g.][]{Proga03, Fukumura10} for newer theoretical models).  Even stronger evidence of magnetic processes driving disk winds and a possible anti-correlation with radio jets has been found in the observations of some stellar-mass black holes \citep{Miller06, Miller08, Kallman09, Luketic10,Blum10}.

Understanding the ultimate influence of these warm outflows on host environments is also interesting and timely.  \cite{Blustin05} considered a sample of Seyfert warm absorbers, and derived typical values for these outflows.  If the winds operate during the entirety of the active Seyfert/AGN phase, then warm absorbers may expel up to $10^{8}~{\rm M}_{\odot}$ of material into their host galaxies.  In large elliptical galaxies, this may be unimportant compared to the stellar mass.  However, this may represent a substantial fraction of the gas in the central spheroid of spiral galaxies.

Estimates of the radii at which warm absorbers exist, and their total mass outflow rates, depend on detailed spectroscopy. Factors such as the gas density, covering factor, and filling factor must be estimated using line ratios and a number of physical assumptions.  Indeed, the {\it geometry} of warm absorber outflows is poorly understood; at present, only lines in NGC 3783 show even modest evidence of P-Cygni profiles \citep{Kaspi02}, likely indicating that the wind is quasi-spherical.  It is notable, then, that the best spectra provide the best evidence that warm absorbers may be disk winds that could be related to fundamental disk physics.

Clearly, deep observations and sensitive spectra are vital to understanding accretion onto, and mass outflows from, the central engine in Seyfert AGN.  To this end, we have analyzed the summed, deep-exposure {\it Chandra}/HETG spectrum of the Seyfert-1 galaxy NGC 4051.  This Seyfert is close ($z = 0.00234$), and the mass of the black hole has been constrained using reverberation mapping techniques (${\rm M}_{BH} = 1.9\pm 0.8 \times 10^{6}~{\rm M}_{\odot}$, \cite{Peterson04}; ${\rm M}_{BH} = 1.7\pm 0.5 \times 10^{6}~{\rm M}_{\odot}$, \cite{Denney09}).  The innermost accretion flow in NGC 4051 has been studied using relativistic disk reflection \citep[e.g.,][or a review, see Miller 2007]{ Ogle04, Ponti06}.  

The X-ray warm absorber in this AGN has previously been studied \citep{Ogle04, Nucita10, Steenbrugge09, Lobban11, Pounds11}. We confirm the multiple absorbing regions with a range of ionization parameters, $1 < \log \xi < 5$ described in these papers. However, we do not find the various high velocity components, i.e. $> 2000$ km s$^{-1}$, claimed by \cite{Steenbrugge09}, \cite{Lobban11}, and \cite{Pounds11}. Furthermore, we present evidence for material arising closer to the central engine than previously claimed, i.e. $< 13000 R_g$. 

Herein, we present phenomenological and self-consistent, physical, photoionization models for the X-ray warm absorber observed in the time-averaged HETG spectrum of NGC 4051.  We find that the wind likely originates close to the black hole, within a few hundred gravitational radii.  We also find modest evidence of P-Cygni profiles in the H-like charge states, suggesting that the hottest, innermost part of the wind is quasi-spherical. A cosmology of H$_0$ = 70 km s$^{-1}$ Mpc$^{-1}$, q0 =0 and $\Lambda_0$ = 0.73 was used throughout this analysis.

\section{Methods}

\subsection{Data Reduction}
We have analyzed twelve \texttt{Chandra} High Energy Grating (HEG) and Medium Energy Grating (MEG) spectra of the Seyfert-1 NGC 4051, between the dates of 09 November 2008 (MJD 54779) and 30 November 2008 (MJD 54800). The event files, redistribution matrices and auxiliary matrices were created via the \texttt{CIAO} commands (version 4.2) \texttt{tgextract}, \texttt{mkgrmf} and \texttt{dmarfadd} respectively. After acquiring the data and response files, we co-added the twelve HEG and MEG observations with the \texttt{HEASOFT}, version 6.9, procedures \texttt{mathpha}, \texttt{addrmf}, and \texttt{addarf}. Although NGC 4051 is know to be highly variable, especially in the X-ray band, these data did not vary in flux by more than a factor of two and in spectral index varies from 1.7 to 2.0. Finally, by utilizing both the positive and negative first order spectra, we were able to get over 0.17 ct/s and avoid pile-up typically associated with the zeroth order spectra. A total exposure time of 308 ksec was achieved.

We utilized the Interactive Spectral Interpretation System (\texttt{ISIS}), version 1.6.1, as well as \texttt{XSPEC}, version 12.6.0, to analyze the time-average spectra employing both phenomenological and photoionization modeling techniques. The data were binned at a minimum of 10 counts per bin.

\subsection{Phenomenological Model}
Initially, we fit the HEG and MEG broadband spectra using \texttt{Xspec} between the ranges of 1.5--20\AA\/ and 1.5--23\AA\/ respectively. The ranges were chosen because the respective sections had the highest sensitivity for each spectrograph, ensuring a maximum signal to noise ratio in each spectrum. The preliminary model consisted of a power-law and a disk blackbody component. The disk blackbody is not likely to be physical; however, it adequately describes the soft-excess seen in NGC 4051 and many Seyfert galaxies \citep[e.g.,][]{Reynolds97, Magdziarz98,Crummy06}. We also include an effective H column density in all models with the component \texttt{phabs}, which is frozen at 1.15 $\times 10^{20}$ cm$^{-2}$ to model the line-of-sight absorption \citep{Kalberla05}. Finally, we included a multiplicative constant, so as to account for any discrepancy between the normalization of the two spectrographs. 

\begin{figure*}[t]
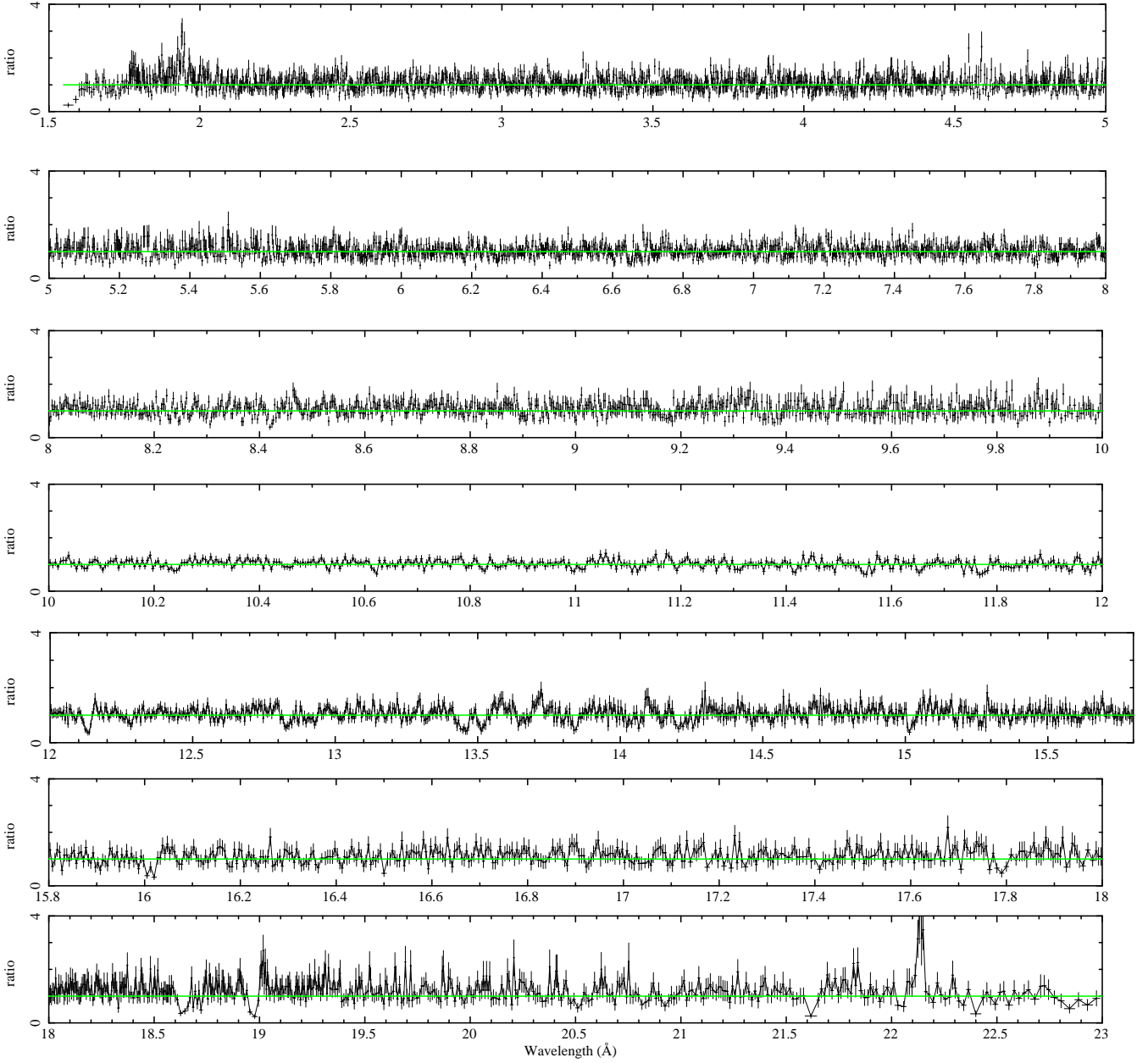

\begin{center}
\includegraphics[height=7in,angle=270]{f1.ps}
\includegraphics[height=7in,angle=270]{f2.ps}
\includegraphics[height=7in,angle=270]{f3.ps}
\includegraphics[height=7in,angle=270]{f4.ps}
\includegraphics[height=7in,angle=270]{f5.ps}
\includegraphics[height=7in,angle=270]{f6.ps}
\includegraphics[height=7in,angle=270]{f7.ps}
\caption{\footnotesize{This figure contains the ratio of the data to the fiducial model with $\chi^2/\nu$ 9607/7516. This model only includes a powerlaw and disk blackbody component as well as an effective H column density and a multiplicative constant, which corrects for the normalization between the HEG and MEG spectra. Clear absorption and emission features can be seen over this simple model. The wavelength plotted is in the observed reference frame. }}
\label{ratiorange}
\end{center}
\end{figure*}

The best-fit parameters of the model with a $\chi^2/\nu$= 9067/7531 are listed in Table \ref{BroadBand}.  The ratio of the data to the continuum model is plotted in Figure \ref{ratiorange}. The power-law index of $\Gamma = 1.94\pm 0.01$ is typical of Seyfert galaxies. The disk blackbody temperature of 0.118$\pm0.005$ keV is also typical of Seyferts \citep[e.g.,][]{Reynolds97}, but again, is not likely to be a physical model.

In general, this continuum-only model is a poor fit to the data, and it is clearly due to the fact that emission and absorption features are present. We note them in the residuals in Figure \ref{ratiorange}. We first employ a phenomenological model as a means of characterizing both of these features. The spectra were fit with \texttt{ISIS} utilizing 2 or 3 \AA\/ sections with a local power-law continuum and a Galactic absorption of 1.15 $\times 10^{20}$ atoms cm$^{-2}$. By allowing the continuum to vary from segment to segment, we were able to precisely detect any underlying features amongst the residuals. We used a Gaussian line profile to describe the residual absorption and emission features. A total of 40 lines were detected and are listed in Tables \ref{lineId} and \ref{lineIdabs}. Of these lines, all but one are detected at least at the 3$\sigma$ confidence level, and the remaining line is detected at a 2$\sigma$ confidence level, but it is expected as part of a pair. The significance of each detection, which is determined using Ftest statistics, is also listed in these tables. The column densities listed in Tables \ref{lineIdabs} are also derived for each ion with known oscillator strength using the relation $W_{\lambda}=(\pi e^2 / m_ec^2) N_j \lambda^2 f_{ij}$, where $W_\lambda$ is the equivalent width, $N_j$ is the column density, $\lambda$ is the rest wavelength, and $f_{ij}$ is the oscillator strength.

We detect a number of different, highly ionized charge states including species of O, Ne, Mg, Si, S and Fe L-shell transitions, as well as narrow fluorescent Fe K$\alpha$ and $\beta$ lines and a broad Fe K$\alpha$ line. The median velocity of all of these features is -580 km s$^{-1}$ , while the mean velocity is -530 km s$^{-1}$ with a standard deviation of 320 km s$^{-1}$ . The emission features have a slightly lower mean velocity of -310 km s$^{-1}$ with a standard deviation of 450 km s$^{-1}$, consistent with emission in the rest frame of NGC 4051. The absorption features have a faster mean outflowing velocity of -630 km s$^{-1}$ with a standard deviation of 200 km s$^{-1}$. The median and mean measurements do not include the  unidentified lines or the fluorescence emission lines.

\begin{figure}[t]
\begin{center}
\includegraphics[height=3in,angle=270]{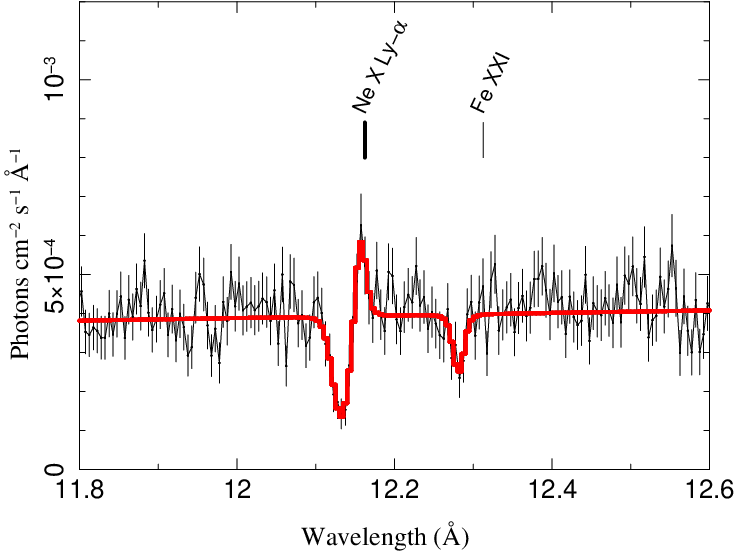}
\caption{\footnotesize{These Ne X and Fe XXI lines are modeled with Gaussian profiles. Here one can see that the absorption is blue shifted with respect to the emission feature seen in the H-Like Lyman $\alpha$ line of Ne X. The P-Cygni-like profile of the H-like line could be real, but may also be a superposition of distinctive wind components. The line markers are in the rest frame of NGC 4051. The wavelength plotted is in the observed reference frame. }}
\label{pcygNe}
\end{center}
\end{figure}

\begin{figure}[t]
\begin{center}
\includegraphics[height=3in,angle=270]{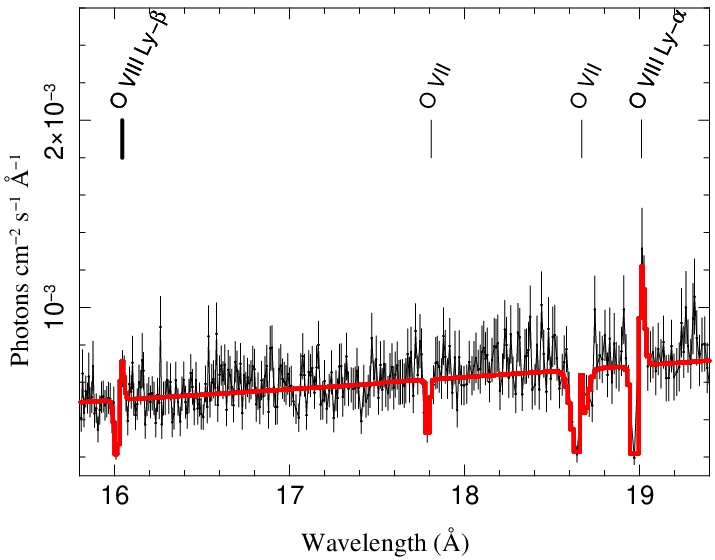}
\caption{\footnotesize{These OVII and OVIII lines are modeled with Gaussian profiles. Here one can see that the absorption is blue shifted with respect to the emission features seen in the H-Like Lyman series of O VIII. The P-Cygni-like profile of the H-like line could be real, but may also be a superposition of distinctive wind components. The line markers are in the rest frame of NGC 4051. The wavelength plotted is in the observed reference frame, and the data was rebinned with a minimum of 30 counts per bin for clarity. }}
\label{pcyg}
\end{center}
\end{figure}

These velocity shifts suggest that the absorption features are produced in a region along our line-of-sight that is moving toward us. Furthermore, the fact that the absorption is blue-shifted with respect to the emission features may suggest the evidence of P-Cygni lines. These features are most clearly seen in Lyman series of both O VIII and Ne X, in particular the lines with rest wavelengths of 18.96 \AA\/, 16.01 \AA\/ and 12.1 \AA\/ (see Figures \ref{pcygNe} and \ref{pcyg}). It should be noted that any emission features in the He-like forbidden or intercombination lines cannot be P-Cygni features, due to the fact that they are optically thin. However, the detection of an absorption feature blue-ward of the emission feature in the H-like lines may give insight to the geometry of this system in the inner highly ionized region.  If these two features are connected and not coincidental, then the lines suggest a spherical shell of material that is out-flowing from the system, and lends itself to determining roughly the covering fraction of this system. However, we cannot rule out that the H-like lines may also be consistent with emission in the rest frame of NGC 4051 and an absorbing region further from the central engine. To date, the only evidence for P-Cygni profiles in Seyfert winds is in NGC 3783 \citep{Kaspi02}.

\subsection{Density Diagnostics}
\label{dens}
\begin{figure}[t]
\begin{center}
\includegraphics[height=3in,angle=270]{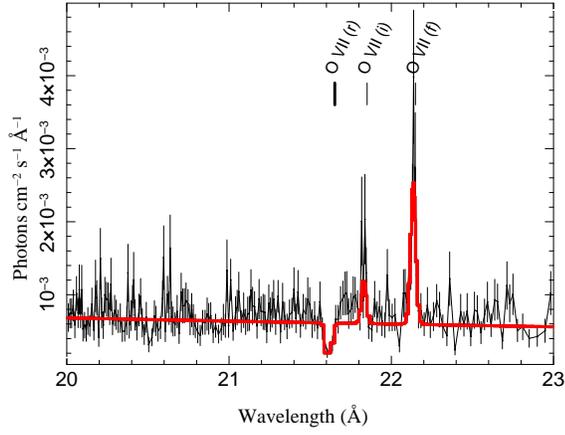}
\caption{\footnotesize{This figure shows the OVII lines characterized with the phenomenological model for both absorption and emission. This He-like triplet gives a density estimate of n$_e\approx$ 10$^{10}$ cm$^{-3}$. The wavelength plotted is in the observed reference frame, while the line identifications are in the rest frame of the host galaxy. }}
\label{OVII}
\end{center}
\end{figure}

There are four particularly relevant sets of line detections found in this data. The first two are the He-like triplet emission lines of O VII and Ne IX, and remaining two are the Fe L-shell doublets, Fe XXI and Fe XXII. The line ratios in the triplets allow for temperature and density diagnostics. However, this assumes that the emission lines originate in the same region as the absorption features and line ratios are not also affected by the ultraviolet radiation field \citep[e.g.,][]{Mauche01}. In addition, in the spectrum of NGC 4051, the emission component of the resonance line (r) is badly blended with the absorption component, as indicated by an absorption equivalent width of O VII 18.65\AA\/ which is larger than the equivalent width of the 21.6\AA\/ line (Figure \ref{OVII}). Therefore, the well-known G ratio is unobtainable, giving only the ratio of the forbidden (f) to intercombination (i) lines.  The ratios
\begin{equation}
R(n_e) = \frac{f}{i}
\end{equation}
\noindent
are 3.4 $\pm$ 1.6 and 1.1 $\pm$ 0.4 for the O VII and Ne IX R ratios. These R ratios are both consistent with the R ratios using the O VII and N VI triplets in work by \cite{Nucita10}. However, \cite{Nucita10} do find a lower limit to the ratio of R $\gtrsim$ 5.6 for the Ne IX triplets which is much higher then our results.
  
The R ratios we present would imply densities of $10^{10}~\rm cm~^{-3}$ and $10^{12}~\rm cm~^{-3}$ for the ratios from O VII and Ne IX, respectively, according to the calculations of \cite{Porquet00} if the UV flux is not important. However, photons near 1620 \AA\/ and 1248 \AA\/ can excite transitions from the metastable levels of O VII and Ne IX, respectively, affecting the ratio just as a high density would.  We do not have simultaneous UV observations, but \cite{Collinge01} report fluxes of about $2 \times 10^{-14}~\rm ergs~cm^{-2}~s^{-1}~\AA ^{-1}$ at both wavelengths. With oscillator strengths and A values from the CHIANTI database \citep{Dere09}, we estimate that the UV excitation would be comparable to the radiative decay rates at distances of $1.8 \times 10^{15}$ cm (7200 R$_g$) and $5.7 \times 10^{14}$ cm (2300 R$_g$) from the central source.

While it is not possible to completely disentangle the density and distance, we can place some limits.  The O VII ratio is consistent with the low density limit.  It implies a density below $4 \times 10^{10}$ cm$^{-3}$ and a distance from the central source above $1.8 \times 10^{15}$ cm (7200 R$_g$) . The Ne IX ratio has larger uncertainties, but taken at face value it would imply a density between $6 \times 10^{11}$ and $3 \times 10^{12}$ $\rm cm^{-3}$ or a distance from the center of about $6 \times 10^{14}$ cm (2400 R$_g$). It seems somewhat unlikely that the Ne IX lines would be formed at a higher density and smaller distance than the O VII lines, and we conclude that one or both of the Ne IX lines is probably affected by the complex Fe absorption structure in that wavelength range.

Additional sets of line detections may suggest a higher density region that is more clearly relevant to the outflow. The previous emission lines were only assumed to originate in the same outflowing region as the absorption features. However, the detection of absorption features that also denote a similar density gives further support to this assumption. These absorption features are the Fe L-Shell density diagnostics of the Fe XXII absorption pair at 11.92 \AA\/ and 11.72 \AA\/, and the Fe XXI absorption pair at 12.285 and 12.38 \AA\/. The Fe XXII pair of lines has a much weaker detection than the triplet emission lines, but we do find the 11.72 \AA\/ line at a $99.997\%$ confidence level, or 4$\sigma$.  The weaker of the two lines, at 11.92 \AA\/, is detected at 2$\sigma$ confidence level, and we do expect this line to be present if the 11.72 \AA\/  line is observed and collisional excitation dominates. The ratio of 11.92 \AA\/ to 11.72 \AA\/ is 0.4 $\pm$ 0.2, and places the ratio in the low density limit. Using the work by \cite{Mauche03} and \cite{Miller08}, we find an upper limit of 7$ \times 10^{12}~\rm~cm~^{-3}$. Furthermore in this density regime of n$_e <7 \times 10^{12}~\rm~cm~^{-3}$, we would also expect the lines from Fe XXI. We do note that the Fe XXI line at 12.285 is detected at a 99.99\% confidence level but not the Fe XXI 12.38. See Figure \ref{pcygNe}. An upper limit to the normalization of the 12.38 is found to be -2.6$\times 10^{-7}$ photons cm$^{-2}$ s$^{-1}$. This places an upper limit to the ratio of 12.38/12.29 of 0.1. Using Figure \ref{chiantife}, this places an even stricter upper limit on this density region of $ n_e < 6 \times 10^{11}$ cm$^{-3}$.

\subsection{The Photoionization Models}
\subsubsection{Parameters}
Because the data are rich with information, we further characterized the X-ray absorption and emission features by implementing more self consistent and physical models. To do so, we used two independent photoionization codes to create these models via XSTAR, version 2.2.0, and Cloudy, version 08.00. These photoionization codes are initially set up with a simple geometry including a spherical cloud, illuminated by a central source, i.e. an AGN. Beyond this, complexity is built into the models by specifying the density, geometry and parameters of the illuminating source.  This technique using multiple photoionization models allows for rigorous tests of the components required by the data. However, a detailed comparison between the two packages is outside the scope of this paper.

We begin with our average incident source, which has a ionizing luminosity of $10^{42}$ ergs s$^{-1}$, consistent with \cite{Maitra11} and \cite{Blustin05}. The incident spectrum has a power-law shape in the X-ray regime with spectral index $\Gamma = 2.0$. Cloudy assumes a more detailed profile in other wavelength regimes, i.e. a blackbody in the UV, while XSTAR only assumes a power-law from 0.1eV to MeV. However, due to lack of contemporaneous data in other wavelength regimes, (e.g. in the UV), a power-law description is adequate for this work. Moreover, \cite{Maitra11} discuss evidence that the weak UV component in NGC 4051 may arise partially from a jet, rather than just the accretion disk. Evidence for jet production in the central region of NGC 4051 is also given in \cite{King11} and \cite{Jones11}.

As the incident spectrum interacts with the absorbing material along our line of sight, the observed features can be produced. We derive the line widths (200  km/s) used in the simulations from the Gaussian fits to the absorption features. The covering fraction of the cloud is approximated at 75\%, taken from the evidence of P-Cygni profiles, (i.e. Figure \ref{pcyg}) and average covering fractions given by \cite{Blustin05} and \cite{Reynolds97}. However, as will be shown below, the covering fraction only impacts the normalization of the emission lines. In addition, the inclination angle derived from the relativistic lines in the subsequent Section \ref{rellines}, suggests that the wind is neither a narrow wind along the polar axis nor a wind along the disk surface. Thus, a large covering factor is a reasonable assumption. A solar metallicity although not derived, was assumed for each model.

Finally the density of each cloud was set according to the tentative density limits found from our line ratios in Section \ref{dens}.  A density of n$_e< 6 \times 10^{11}$ from Section \ref{dens} used as a constraint on the initial set of models, which were created with densities of 10$^{10}$ and 10$^{11}$ cm$^{-3}$. The density of these regions is held constant, and we find that both density regions are required by the data. The 10$^{11}$ cm$^{-3}$  characterizes the more highly ionized regions, based on the conservative Fe XXII and Ne IX constraints, and the 10$^{10}$ cm$^{-3}$ delineates the less ionized components at larger radii, based on the O VII triplet diagnostics. 

The data require three regions for both XSTAR and Cloudy models. The absorption models were included as multiplicative 3-dimensional tables varying the ionization parameter, column density and redshift for XSTAR and inner radius, column density and redshift for Cloudy. The emission models were included as additive 4-dimensional tables, which varied the above 3 parameters as well as a normalization parameter. We linked the column density and ionization parameter for the XSTAR absorption and emission components, and the column density and inner radius for the Cloudy absorption and emission components. Thus, the redshift and normalization were allowed to vary in the emission components. To compare between the parameters of the two models, we use the relation,
\begin{equation}
\label{xi}
\xi = \frac{L_{ion}}{n(r) r^2}
\end{equation}
where $\xi$ is the ionization parameter, $L_{ion}$ is the ionizing luminosity, n(r) is the electron density and r is the radius. 

\subsubsection{Results}
\label{res}
Various combinations of the different density regions were tried, but inevitably two faster outflowing, highly ionized regions ($\log\xi \approx 3.3-5.0$) and a slower, less ionized region ($\log\xi \approx 1-2$) were always present when globally fitting the spectra for XSTAR and Cloudy. The best fit was in fact given by three absorption regions, two at 10$^{11}$ and one at 10$^{10}$ cm$^{-3}$ with a $\chi^2/\nu = 8177/7516$ and $\chi^2/\nu = 8713/7516$ for the XSTAR and Cloudy models, respectively (see Table \ref{xstar} and \ref{cloudy}). For completeness, we found a formally worse fit with three absorbing regions, at a density of 10$^{10}$ cm$^{-3}$,  with a $\chi^2/\nu = 8205/7516$ for the XSTAR model. Assuming this is a change of two degrees of freedom, i.e. the density, and using an Ftest, we find that the XSTAR model with 10$^{11}$ cm$^{-3}$ components is a 4$\sigma$ improvement over models with 10$^{10}$ cm$^{-3}$ components. Repeating this with the Cloudy models, we found a marginally better fit with three 10$^{10}$ cm$^{-3}$ at $\chi^2/\nu = 8709/7516$. However, this is not statistically significant at only a confidence level of 1.2 $\sigma$. We decided to include the Cloudy models with two 10$^{11}$ cm$^{-3}$ and one 10$^{10}$ cm$^{-3}$ density components for like comparison with the XSTAR model. (See Figures \ref{xstarrange} and \ref{cloudyrange}). 

In addition to using the Ftest to justify our use of the higher density models, we also refit the spectra with and without the Fe XXII and Fe XXI lines in an effort to assess the extent to which these density sensitive lines were driving the fits. This was done only in the XSTAR models. By removing 0.3 keV regions around the Fe diagnostic lines, and refitting the data, we find the ionization parameters of all three components decrease from $\log(\xi)$ = 4.5, 3.3 and 1.0 to $\log(\xi)$= 3.6, 2.6 and 0.8. However, if we only include the regions around the Fe diagnostic regions in the spectral fit, the ionization parameters of the two highest regions stay roughly the same, while the lowest ionization component increases; $\log(\xi)$ = 4.5, 3.6 and 2.4. This is demonstrating that the Fe L-Shell diagnostic lines are in fact driving the spectral fits, for when they are included the ionization stays the roughly same. Conversely, when the lines are excluded, the ionization of all three components tends to decrease to lower ionizations. It therefore seems a reasonable assumption to use the higher density models. We have repeated this process assuming all the components had a lower density of 10$^{10}$ cm$^{-3}$, and find the same trend in the ionization parameters but a statistically worse fit for each scenario.

As a caveat, one should be careful interpreting these results, as the densities are upper limits. Reducing the density will tend to increase the radius at which these features are detected.

Besides density, we also explored the effects of velocity broadening and covering fraction. In order to do so, we varied these parameters in the best-fit XSTAR model. Due to the resolution of the HETG spectrograph, we do not expect to resolve lines widths less than 150 km/s. However, larger velocity widths can be resolved. We fit the data with an XSTAR model that has a velocity broadening of 300 km/s and find a worse fit at $\chi^2/\nu$ = 8275/7516. Finally, we also verified our choice of covering fraction. Reducing the covering fraction to only 25$\%$ in the XSTAR model results in a worse fit of $\chi^2/\nu$ = 8188/7516. Although the 75\% covering fraction model is a 3 $\sigma$ improvement over the 25\% model, the change in parameters is only in the normalization of the emission lines. 

In general, the absorption regions specified by the fits are quite similar between the two packages. The Cloudy models are typically outflowing at a higher velocity than the equivalent XSTAR component, but in both the XSTAR and Cloudy models, the emission features are consistent with being redshifted with respect to their  absorption counterparts. Furthermore, in both the XSTAR and Cloudy models, the highly ionized region of $\log \xi \approx 3.3-3.6$, predominately characterizes the higher energy absorption features. This includes the lines especially below 14\AA\/ such as the He-like Si XII and Mg XI, the H-like Mg XII, and the Lyman Series of Ne X. Conversely, the lowest ionization region predominately characterizes the lower energy absorption features. This includes the He-like O VII lines and H-like O VIII lines above 16 \AA\/. This $\log \xi \approx 1-2$ also characterizes the O V feature at 22.4 \AA\/. Additionally, \cite{Collinge01} also find a C IV absorption feature at 1548.2 \AA\/ and 1550.8 \AA\/, which is formed at $\log \xi \approx 1.4$. 

Of the absorbing regions, the most highly ionized component is the least constrained for it reaches the maximum ionization parameter in both of the XSTAR and Cloudy grids. Thus, the $\xi$ is a lower limit in this model. However this component is required by an F--test and is required to characterize the highly ionized Fe L--Shell lines including the density dependent, Fe XXII. The two highest ionization parameters in both models, and therefore smallest inner radii, place the absorbing regions within a few hundred gravitational radii of the SMBH assuming the density is constant with radius. At this radius one would expect line broadening to be approximately the orbital speed, resulting in a broadening of $\approx$ 4$\times 10^{4}$ km s$^{-1}$. Clearly this is at least two orders of magnitude higher than the observed 190 $\pm$ 150 km s$^{-1}$ line widths derived from our Gaussian fits.  However, due to the geometrical effects of encoding orbital motion into the absorbing region because of a small illuminating source assumed to be at $<$10 R$_g$ and a distance of 500 R$_g$, the resulting broadening would only encode 2\% of the orbital motion, i.e. $\approx$ 300 kms$^{-1}$. This is consistent with the 1$\sigma$ error of our measurements. Alternatively, if the density of these regions is not constant but scales as r$^{-2}$, as thought to be typical of winds \citep[e.g.,][]{Miller08}, or even r$^{-1}$, then the inner radius need not be as close to the SMBH and can also remedy this discrepancy.

Where the XSTAR and Cloudy models differ seems to be predominantly in the range of 10 -- 16 \AA\/, and they differ mainly in the $\log \xi \approx 1-2$ components.   As seen in Figures \ref{xstarrange} and \ref{cloudyrange}, one can readily notice the differences in the absorption features especially around 14\AA\/, which is likely the Fe L-Shell charge states. The absorption discrepancy between the two models is likely to be explained by the inclusion of the Fe L-shell transitions of Fe XVII -- Fe XXII. The Fe L-Shell transition probabilities, ionization rates and cross-sections are less well known than the H-like and He-like transitions of O or Ne. So it should not be surprising the two models differ. The exact differences between the models are outside the scope of this paper, but we note that in a general sense the models follow the same trends.

\subsubsection{ Fluorescence Fe K$\alpha$ and O VIII Lyman $\alpha$}
\label{rellines}
Also interesting in these fits are the lack of predicted Fe K$\alpha$ or $\beta$ emission at 1.94 and 1.75 $\AA\/$ respectively. Cloudy fits the weak Fe XXIV emission line at 1.85\AA\/ but not the fluorescence lines at either 1.77 or 1.95\AA\/, and XSTAR only fits the slightest of emission at the Fe K$\alpha$ wavelength. Conversely, both these lines as well as a broad component at a minimum of a 99.5\% confidence level are required by the phenomenological fits. This is likely due to the fact that these fluorescence lines arise from neutral or very low ionization states of iron, and would not be modeled by the high ionization parameters in our models. We argue that these lines come from a different region entirely because of their detection in the phenomenological model and lack thereof in the photoionization models as well as the ionized states. The narrow lines are the reflection Fe K $\alpha$ and $\beta$ lines probably produced in the outer regions of the accretion disk. Conversely, the broad component is thought to originate within a few hundred gravitational radii of the black hole. This is consistent with what \cite{Young05} find while studying MCG -6--30--15. They also state that the curvature produced by photoionization models could in principle produce a redwing effect in the Fe K$\alpha $ regions, but we do not see this effect. 

As a result, we decided to model the Fe K$\alpha$ and $\beta$ lines using two narrow Gaussian components fit to the data at 6.41 and 6.97 keV in the rest frame of the galaxy to describe the narrow components as well as \texttt{Kerrdisk} to model the broad component. We also include the XSTAR models.  \texttt{Kerrdisk} is a ray tracing model that accounts for both relativistic beaming and light bending effects in the vicinity of the accretion disk and black hole \citep{Brenneman06}. In addition, it also assumes material within the radius of marginal stability is very highly ionized and has a low optical depth, thus lowering its contributions to the line profile. Finally, it further excels over other models, for not only does it allow the inner and outer radii, inclination angle, and emissivity index of both the inner and outer disk to vary, but it allows the spin parameter to vary. 

In our model, we fixed several of the parameters of the \texttt{Kerrdisk} model including the energy of the line to 6.39 keV in the rest frame of NGC 4051, the emissivity index of the disk to be 3, the inner radius to be 1 R$_{ms}$, and the outer radius to be 400 R$_{ms}$, where R$_{ms}$ is the radius of marginal stability. We find that the best fit values using \texttt{Kerrdisk} at a $\chi^2/\nu$ = 8085/7507 and the model is plotted in Figure \ref{fefl}.  We find a normalization of 4.0 $\pm$ 0.6 $\times 10^{-5}$ photons cm$^{-2}$ s$^{-1}$, and a maximum spin of a=0.998$_{-0.088}$ with an inclination angle of 31.6$^{+1.1}_{-1.7}$$^\circ$. A maximally spinning black hole is in agreement with the results of \cite{Ogle04} and the inner radius is in agreement with \cite{Steenbrugge09}. However, the inclination we find is lower and inconsistent with the 48$^{\circ}$ that both report.     

In addition to the Fe K$\alpha$ emission line, we also explored the possibility of a relativistic O VIII Lyman $\alpha$ line found by both \cite{Ogle04} and \cite{Steenbrugge09}. Evidence of such a line would suggest an over-abundance of Oxygen, which may be similar to the over-abundance of Iron seen in Seyfert-1, 1H 0707-495 \citep{Fabian09}. We include this line with \texttt{Kerrdisk} fixing the rest frame energy at 0.654 keV ($\chi^2 / \nu$ = 8169/7513). This component was required at a confidence level of just over 2$\sigma$, at 97.3\%. The resulting parameters of the best fit model include a spin of a=0.998 (unconstrained) with an inclination of 37.6$^\circ$$^{+4}_{-1.4}$ and a normalization of 1.4 $_ {-0.3}^{+0.7}$ $\times 10^{-4}$ photons cm$^{-2}$ s$^{-1}$. Again we find a maximally spinning black hole, although not constrained, consistent with the aforementioned Fe K$\alpha$ line parameters. The inclination angle has increased but is not consistent with \cite{Steenbrugge09} and \cite{Ogle04}.

If we tie the spin parameter, inclination angle and inner radius of the O VIII Lyman $\alpha$ line to that of the Fe K $\alpha$ and then refit the data, we find a similar significant detection at just over $2\sigma$, 96.9\% ($\chi^2 / \nu = 8080/7506$) . However, the Fe K$\alpha$ line does drive the spin to 0.998$_{-0.08}$ with an inclination angle of 31.7$^{+0.8}_{-0.7}$$^\circ$. The normalization of the O VIII Lyman $\alpha$ line is 1.2 $\pm$ 0.5 $\times 10^{-4}$ photons cm$^{-2}$ s$^{-1}$, and the normalization for the Fe K $\alpha$ line is 4.0 $\pm 0.6$ $\times 10^{-5}$ photons cm$^{-2}$ s$^{-1}$. This model, along with the aforementioned models is plotted in Figure \ref{okerr}.

\section{Discussion}
\subsection{Physical Characteristics}
Finding three distinct ionizing regions illuminates different attributes of the winds in NGC 4051. We are able to constrain where these components are launched, the size of the emitting regions, how ionized the regions are, and the total amount of material that is observed to leave the system. 

Table \ref{photo} lists the aforementioned ionization parameters and radii of each of the three regions determined via XSTAR and Cloudy. \cite{Denney09} use reverberation mapping techniques to place the Broad Line Region (BLR) at 1.87 $\pm$ 0.5 light days or 4.84 $\pm$ 1.4 $\times 10^{15}$ cm. The lowest XSTAR ionization component is consistent with this distance, while all the other components are interior to the BLR. This proximity to the SMBH suggests that the accretion disk may play a role in the launching mechanisms. 

Also listed in Table \ref{photo} is the radial extent of each region. This is estimated by using the simple approximation $\Delta(r) = N_H/n$. The size of each region is between 2 $\times 10^{9}$ and $2.2 \times 10^{10}$ cm, less than 1\% of the radial distance to the source.

 Because each region has a relatively small radial extent, the mass outflow rate is low. Following the work of \cite{Blustin05}, the mass outflow rate is determined assuming a modified spherical outflow. As \cite{Blustin05} describe, the outflow is most likely in clouds and filaments, so caution should be taken when interpreting these results. The outflow rate is described as follows,
\begin{align}
\label{Mout}
\dot{M}_{out}= & r^2 \rho (r) v(r) \Omega \approx 1.23 r^2 m_p \bar{n}(r) v \Omega 
\end{align}
Here we assume solar abundances, $\rho$= 1.23 m$_p\times \bar{n}(r)$, where m$_p$ is the mass of a proton, and a covering fraction of $\Omega$= 0.75 $\times$ 4$\pi$.  It should be noted, that the density is assumed to drop off as 1/r$^2$, so n(r)=n(R)R$^2$/r$^2$. To relate the average number density, $\bar{n}(r)$  to the measured number density, $n(r)$, a volume filling factor, C$_v(r)$, needs to be included as so,
\begin{align}
\label{n}
\bar{n}(R) = & n(R) C_v(R) 
\end{align}
A volume filling factor C$_v$ is derived for each case using the following form,
\begin{equation}
\label{cv}
C_v = \frac {\xi N_H R}{L_{ion}}
\end{equation}
where $\xi$ is the ionization parameter, N$_H$ is the column density, R is the distance from the source to the absorbing region and L$_{ion}$ is the ionizing luminosity. These values are given in Table \ref{photo}.

Finally, combining equations \ref{xi}, \ref{Mout}, and \ref{n} one gets,
\begin{align}
\dot{M}_{out} \approx  & \frac{1.23 m_p L_{ion} C_v(R) v \Omega}{\xi}
\end{align}
The resulting mass outflow rates are given in Table \ref{photo}. The total mass outflow predicted by XSTAR is 7.9$^{+ 9.6}_{-7.4} \times 10^{-6}$ M$_\odot$yr$^{-1}$. Cloudy models give a higher predicted outflow rate of 1.3$^{+0.8}_{-0.9} \times 10^{-5}$ M$_\odot$yr$^{-1}$. In both models, the less ionized and slower moving component dominates this result by almost an order of magnitude. This can be attributed to the large covering factor assumed to be 75\% and the large radius at which it is observed. The outflow rates may be lower limits, if gas flows at smaller radii with a very high ionization that inhibits its detection using atomic features.

The observed  mass outflow rate is very low compared to the implied mass accretion rate onto the SMBH. The mass accretion rate is, $\dot{M}_{acc} =  L_{bol} / {(c^2 \eta)} $, where L$_{bol}$ is the bolometric luminosity of the NGC 4051\citep[L$_{bol}$=10$^{43.4}$ ergs s$^{-1}$,][]{Blustin05}, and $\eta$ is the efficiency factor, typically assumed to be 10\%. With $\dot{M}_{acc}$= 0.0047 $M_\odot yr^{-1}$, one finds that $\dot{M}_{out}^{total}/ \dot{M}_{acc} $ is approximately 0.002 and 0.003 for XSTAR and Cloudy respectively. For reference, the Eddington accretion rate would be $\dot{M}_{Edd} \approx$ 0.037 $M_\odot yr^{-1}$, a much higher rate than either mass accretion rate or the outflow rate. The low mass outflow rate can be attributed to the very small filling factors derived from equation \ref{cv}. \cite{Blustin05} also find a small mass outflow rate of 0.0008 M$_\odot$yr$^{-1}$, but it is still an order of magnitude larger than the mass outflow rates found in this paper.

\subsection{Physical Scenario}

In addition to the amount of material that is being driven from the central regions, we can also comment on the physical nature of this system. In previous work, authors have claimed evidence for both shocks, and two phase regions \citep[i.e.,][]{Pounds11,Krongold07}. Here we discuss and compare our work to previous analysis of the warm absorbing winds in NGC 4051.

The three regions in both of the XSTAR and Cloudy models follow the trend where the most ionized regions are closest to the source and fastest moving. This is in contrast to the work done by \cite{Nucita10} who find only one absorption component, which is moderately ionized and not outflowing. The lack of blue shift may be result of the lower resolution of the RGS spectra. In addition, \cite{Nucita10} also find two emission components, one of which is consistent with our $\log \xi = 3.3$.

On the other hand, \cite{Lobban11} and \cite{Pounds11} find a similar trend to our work, where the outflowing velocity decreases with decreasing ionization parameter. In particular, \cite{Lobban11} find four zones with outflowing velocities ranging from -180 to - 820 km s$^{-1}$ and ionization parameters ranging from $\log \xi$ = -0.86 to 2.97. They also find an additional highly ionized region, $\log \xi = 4.1$ which is outflowing very quickly at 5800 km s$^{-1}$.  We do not find the lowest outflowing nor the lowest ionization regions, but our data are consistent with all but the largest velocities, i.e. 400 $< v_{out}<$ 710 km s$^{-1}$. This is likely due to the fact that the the HETGS is less sensitive to the low ionization features. Because there is a trend between increasing velocity and ionization, we would not expect to detect these slow moving, low ionization features. Table \ref{compare} lists these parameters for comparison.

\cite{Pounds11} find five absorption regions with velocities ranging from +120 to -5880 km s$^{-1}$ with ionization parameters spanning $\log \xi$ = 0.32 to 2.97. Again, we do not find the lowest velocity component nor the two highest velocity components, which are outflowing faster than 3800 km s$^{-1}$ (See Table \ref{compare}). Several of their high velocity components are outside our wavelength range, but a majority of these highly ionized, highly velocity components are within our range. We note that the components b and c seen Figure 2 of \cite{Pounds11} are also readily explained by an O VII line rather than a highly blue-shifted O VIII Lyman $\alpha$ line, as shown in our XSTAR and Cloudy models (see Figures \ref{xstarrange} and \ref{cloudyrange}). Additionally, \cite{Pounds11} note that the high velocity components they observe with the EPIC CCD cameras include Mg XII through Ar XVIII. We do not find these in our high resolution spectra.

The multiple regions described in \cite{Lobban11} and \cite{Pounds11} are both discussed as a continuous region that spans a variety of different ionization states in the vicinity of a shocked shell. The authors suggest that their multiple regions are consistent with a constant mass outflow rate, i.e. $\dot{M}_{out} \propto \frac{L_{ion}  v}{\xi} = $ constant. Figure 8 in \cite{Pounds11} demonstrates the necessary linear dependence of the observed ionization parameter and velocity. However, we do not find a similar linear dependence between the velocity and ionization parameters. Our ionization parameters span four orders of magnitude while the velocity shifts only span one.

\cite{Steenbrugge09} also find a number of ionization regions  in their a variability study of NGC 4051 using \textit{Chandra} LETGS. In their highest signal to noise spectrum, C, they note four regions with ionization parameters ranging from $\log \xi$ = 0.07 to 3.19. Instead of evidence for a shocked flow, \cite{Steenbrugge09} state that the ionized regions must originate within a 0.02 -- 1 pc distance of the central engine, i.e. much closer than \cite{Pounds11} and \cite{Lobban11}. This distance is more consistent with our results but still formally discrepant. Further, \cite{Steenbrugge09} do note a lack of variation in three ionizing absorbers, placing a lower limit on the distance to the source, we suggest that there is a different scenario that could account for the lack of variability. If there is a continuous, flat distribution of absorbing column density as a function of ionization parameter, a change in the ionizing flux could shift the ionization parameters in such a way that there would be little change in the absorption lines; likewise, one would not see the associated variability following a change in the continuum. This allows the material to lie closer than previously derived from reionization timescales. We note that one component does show slight variability, component 2, which shows a 2.3$\sigma$ drop in ionization parameter over the four epochs. This is discrepant with this scenario. However, \cite{Steenbrugge09}  note that they freeze the column density for the spectra with low signal to noise, which could influence the ionization parameters because the two parameters are not clearly independent of one another. This may influence the aforementioned variability.

In addition, \cite{Steenbrugge09} find a range of velocities for their absorbing regions, from -210 to -4670 km s$^{-1}$. Again, we do not find that our velocities are in agreement with the highest outflows. This is the result that our highest ionizing component, $\log \xi >4.5$ characterizes the Fe L-shell transitions, instead of the C VI and N VII Lyman $\alpha$ lines. The velocity as well as ionization parameters also given in Table \ref{compare} for easy comparison. 

Conversely, \cite{Krongold07} find a two phase medium which is not consistent with the shocks described by \cite{Pounds11} and \cite{Lobban11} or the multiple regions given by \cite{Steenbrugge09}. Instead, the authors conclude that the two absorbing regions are in pressure equilibrium with a pressure on order of 10$^{12}$ K cm$^{3}$. A constant pressure would conflict with a constant mass outflow rate. In our work, we constrain the temperature to be 5$\times 10^{5}$ K in XSTAR, forcing the two highly ionized regions to be in pressure equilibrium, and the third region not in equilibrium. Cloudy finds the equilibrium temperature which varies slightly between the models resulting in temperatures of approximately 4$\times 10^{4}$, 3$\times 10^{5}$ and 7$\times 10 ^{5}$ K for the components of $\log n_e$ = 10, 11 and 11 respectively. Because the Cloudy fit is poor, there is only suggestive evidence that the components are not in pressure equilibrium. We would expect the temperature to increase with decreasing density for a constant pressure.

\cite{Krongold07} also find a lower electron number density than we do, by 3-4 orders of magnitude, using variability analysis of an approximately 50 ksec \textit{XMM-Newton} RGS spectrum. They also find higher ionization parameters than we find, and their photoionization model fails to characterize particular absorption features, most notably the Fe XVII line at 15 \AA\/. (See Table \ref{compare}) This discrepancy may be the result of the \textit{Chandra} HETG spectra used in our work, which affords us a better sensitivity. However, we would like to note a similar trend between our work and \cite{Krongold07}, in that we find at least two regions with ionization parameters that differ by more than two orders of magnitude.

In general we can make these conclusions following a detailed comparison to the literature,
\begin{enumerate}
\item NGC 4051 is highly variable and discrepancies between different observations are to be expected. Moreover, the different resolutions and wavelength ranges of XMM and Chandra will emphasize different features.
\item We find a low ionization component between $\log \xi = 1-3$ that is blue-shifted at rate of 200--600 km s$^{-1}$ consistent with other work on NGC 4051 \citep[e.g.,][]{Krongold07,Steenbrugge09,Lobban11,Pounds11}.
\item We do not find components $\log \xi < 1$, unlike \cite{Steenbrugge09}, \cite{Lobban11}, and \cite{Pounds11} . 
\item We do not find the high velocity components, i.e. $> 2000$ km s$^{-1}$ \citep{Collinge01,Steenbrugge09,Lobban11,Pounds11} . These high velocity, highly ionized components characterize C VI and N VII Lyman $\alpha$ lines \citep{Steenbrugge09} as well the O VIII Lyman $\alpha$ lines \citep{Pounds11} and the Fe XXV and Fe XXVI lines at 6.8 and 7.1 keV respectively \citep{Lobban11}. The velocity shifts that these authors find, although high, are inconsistent with one another (See Table \ref{compare}). Our work shows the highest ionization parameters are required by the Fe L-shell charge states but not at such velocities. Furthermore, the high velocity component c in Figure 2 of \cite{Pounds11}, the O VIII Lyman $\alpha$ line can also be interpreted as the O VII 18.627 \AA\/ absorption line.  The 1s2 - 1s3p line at 18.6 A is expected to be $1/4$ as strong as the 21.6 \AA\/ 1s2--1s2p line, but \cite{Pounds11} Figures 2 \& 3 show it to be approximately the same strength. We interpret this as partial filling of the 21.6 \AA\/ absorption feature by the resonance line emission that must be present along with the observed forbidden and intercombination lines. However, this could also be due saturation of the 21.6 \AA\/ line.
\item \cite{Steenbrugge09} discuss lower limits to radii due to lack of variability in the ionization parameters; we suggest that a continuous set of ionization parameters could also account for the lack of variability at a smaller radii. 
\item Following \cite{Ogle04} and \cite{Steenbrugge09}, we can also report the possibility of an O VII Lyman $\alpha$ relativistic line.  In addition, we find the data require a Fe K$\alpha$ relativistic line. Taken literally, the fits suggest a maximally spinning black hole, a=0.998$_{-0.08}$ with an inclination angle to 31.7$^{+0.8}_{-0.7}$$^\circ$, when these parameters are linked between these two relativistic lines. 

\end{enumerate}

Our analysis shows characteristic traits in common with both the two phase medium, i.e., at least two components that span a wide range of ionization parameters, and the continuous range in ionization parameters and velocities. However, neither of these scenarios are entirely consistent with the data treated in this paper. As a result, we believe that we may be seeing at least two regions separated by a large distance. The inner--most absorbing region is within a few hundred R$_g$ and is made up of at least two highly ionized components, i.e. $\log \xi > 3.3.$ which may be in pressure equilibrium. In fact, there may even be a range of ionization parameters in this region. The other component which is further out is also outflowing, but may not be related to the inner part of the accretion flow. The outer component is not in pressure equilibrium and could be slowing down due to interactions with the host galaxy. It is interesting to note that \cite{Detmers11} describe in Seyfert Mrk 509, a two-component absorbing region with high ionization parameters, and a tentative third lower ionization region. Similar to this work, \cite{Detmers11} describe two discrete components which show a range in ionization states within each component but are not continuous between the two. In addition, the two faster outflowing components at -319 and -770 km s$^{-1}$ found in Mrk 509 are also similar to our work \citep{Detmers11}.

\subsection{Launching Mechanisms}
Regardless of the exact geometry, it is quite intriguing that both of the photoionization codes place a portion of the absorbing material very close to the central engine, suggesting that it is launched in the vicinity of the accretion disk. At such a small radii, thermal pressure is ruled out as a mechanism for driving the winds. The escape temperature at a few hundred gravitational radii is greater than 10$^{11}$ K, while the gas temperature in our models is only on order of $5\times 10^{5}$ K making escape due to thermal motions quite difficult. Consequently, either radiation pressure or magnetic fields are needed to drive the winds. To be driven by radiation pressure, the winds need to have a strong source of UV radiation as well as a large cross section to interact with the UV photons. However, each of these absorbing regions has a ionization above $\log \xi > 3.3$, which lowers the radiation pressure multiplier to very close to 1 \citep{Kallman82}. Therefore, the gas does not gain any significant additional acceleration from radiation pressure at these high ionization states \citep[e.g.,][]{Proga00}. 

This leaves the most probable launching mechanism to be magnetic fields. One particular scenario is described by \cite{Blandford82} who suggest that the winds can be launched from magnetic field lines that are tied to the accretion disk. The ionized gas flows along the magnetic field lines, and due to the rotation of the disk, centrifugal forces launch the ionized material. In this picture, the wind would carry away most of the angular momentum and much of the energy of the accreting gas. Therefore, the standard estimates for mass accretion rate based on luminosities would be an underestimate. Another way to launch the winds via magnetic fields is through magnetic pressure generated through the magnetorotational instabilities (MRI). \cite{Proga03} predict this type of wind to generally have high density and low velocity, which is in accordance with our observations. One of the main differences in these two models is that the centrifugal winds are launched vertically where as the magnetic pressure winds are launched torodially. Unfortunately, we do not have enough sensitivity to measure any rotational velocity encoded into the absorption features to distinguish between the two. 

\cite{Behar09} also finds evidence for magnetically driven Seyfert outflows, which lends further support to our observations. He performs a study using the absorption measure distribution (AMD) as compared to the ionization parameter, $\log \xi$, of 5 Seyferts. In general, he finds a very shallow relation between AMD and $\log \xi$. This can be interpreted as consistent with non-spherical MHD outflows, assuming the variations in $\log \xi$ are not due to a turbulent interstellar medium. The resulting density profile scales inversely with radius, n $\propto$ r$^{-1}$. 

\section{Conclusion}
We have used two separate modeling techniques, a phenomenological model and photoionization models, to quantify the absorption and emission features seen in the \textit{Chandra} HETG spectra of NGC 4051. Amongst the results are three sets of line detections that afford us the ability to diagnose the density of the material. These are the He-like Ne IX and O VII triplets as well as the tentative Fe L-shell doublet, Fe XXII. We also constrain an upper limit to the Fe XXI ratio, another Fe L-shell density diagnostic. In particular, the fact that the Fe XXI and Fe XXII ratios are absorption features means that they are particularly effective probes of the density of the absorbing regions in so much that it is created within the absorbing region itself, whereas the emission features may come from another nearby region.

Following this phenomenological study, we implemented more self-consistent photoionization models via XSTAR and Cloudy. We used electron densities of n$_e=10^{10}$ and n$_e=10^{11}$ cm$^{-3}$ based on the gas density diagnostics, and the data then require three different absorption regions. The two highest ionized regions with  $\log \xi > 3.3$, are at the higher density, while the lower density region is less ionized, with $\log \xi \approx 1-2$.  All three components may originate just at or within the BLR. The highly ionized regions likely originate close to the black hole, i.e. less than $300 R_g$ (assuming no radial drop-off in density) and have a large velocity as compared to the lower ionization region that has a lower velocity and resides farther out, i.e. greater than $3300 R_g$. Velocity broadening of the inner-most components would nominally show larger absorption line widths. However, in Section \ref{res} we demonstrate that a small central source \citep[$<$10 R$_g$][]{Chartas09} may only impart a small portion of the rotational velocity in the absorption components.

Furthermore, we demonstrate that the mass outflow rate is small compared to the overall mass accretion onto the black hole. It is possible that magnetic fields are partly responsible for driving these winds, at least in the regions closest to the black hole. This is very similar to the galactic black holes, GRO 1655-40 \citep{Miller08}, H1743-322 \citep{Miller06b}, 4U 1630-472 \citep{Kubota07} the dwarf nova, OY Carinae \citep{Mauche00} and the Seyfert, NGC 4151 \citep{Kraemer05},  in which magnetic fields are also found to potentially be important in driving outflows.

Finally, we note the data require a relativistic Fe K$\alpha$ line, which nominally implies a maximal spin 0.998$_{-0.08}$ with an inclination angle of 31.7$^{+0.8}_{-0.7}$$^\circ$. At a 2 $\sigma$ level, the data require a relativistic O VIII Lyman $\alpha$ line. \cite{Steenbrugge09} and \cite{Ogle04} also find evidence for a relativistic O VIII Lyman $\alpha$ line.

\vspace{0.2in}
The authors would like to thank the anonymous referee for their invaluable comments and improvements to this paper.

%-------------------------------------Table Start---------------------------------------------------------
\begin{deluxetable*}{c c c c c }
\tablecolumns{5}
\tablewidth{0pc}
\tablecaption{Broad-Band Parameters}
\tablehead{ \colhead{$\Gamma$} & \colhead{Normalization} & \colhead{Constant} & \colhead{Diskbb}  & \colhead{Normalizatoin}  \\  & (10$^{-3}$ photons cm$^{-2}$ s$^{-1}$ keV$^{-1}$) &  &(keV) & $\left( \left( \frac{R_{in}/km}{D/kpc} \right)^2\cos \theta \right)$}
\startdata
 1.94 $\pm$ 0.01 & 4.24 $\pm$ 0.03 & 1.062 $\pm$ 0.005 &  0.118 $\pm$ 0.005 & 6610 $_{-910}^{+1100} $ \\
  \enddata
\label{BroadBand}
\tablecomments{This table gives the parameters for the broad-band continuum of both the HEG and MEG spectra. Because of a soft excess component in the MEG spectra, a disk blackbody component was included. The disk blackbody component should not be regarded as physical. The multiplicative constant is applied to the MEG spectrum to correct for the slight discrepancy between the flux in the two spectrographs. Galactic absorption of 1.15 $\times 10^{20}$ atoms cm$^{-2}$ was also included. $\chi^2/\nu$ = 9067/7531}
\end{deluxetable*}

%-------------------------------------Table End---------------------------------------------------------

%-------------------------------------Table Start---------------------------------------------------------
\begin{deluxetable*}{l l l l l l l l }
\tablecolumns{8}
\tabletypesize{\scriptsize}
\tablewidth{0pc}
\tablecaption{Emission Line Parameters}
\tablehead{ \colhead{line center} & \colhead{$\sigma $} & \colhead{Normalization} & \colhead{EW} & \colhead{$\lambda_\circ$ }& \colhead{velocity } & \colhead{Ftest} & \colhead{ID} \\ (\AA\/) & (m\AA\/) & (10$^{-5}$ ph cm$^{-2}$ s$^{-1}$)  & (m\AA\/) & (\AA\/) & (km s$^{-1}$) & &}
\startdata

 22.135 $\pm$ 0.003  &    12.8 $_{-    4.6}^{+    4.9}$ &    8.73 $_{-    1.37}^{+    1.40}$ &  132 $\pm 21$ & 22.101 &   -240 $_{-  50}^{+  40}$ & 5.97$\times 10^{-08}$  & O VII \\
  21.831 $_{-0.027}^{+0.026}$ &    11.6 $_{-    5.6}^{+   10.3}$ &    2.54 $_{-    0.88}^{+    1.30}$ &   37.8 $_{-   13.1}^{+   19.4}$ & 21.807 &   -370 $_{- 370}^{+ 360}$ & 5.37$\times 10^{-02}$ &    O VII \\
  19.013 $_{-0.003}^{+0.006}$ &    12.4 $_{-    7.4}^{+    6.5}$ &    2.52 $_{-    0.92}^{+    0.79}$ &   37.6 $_{-   13.7}^{+   11.9}$ & 18.967 &     18 $_{-  40}^{+  99}$ & 2.10$\times 10^{-04}$& O VIII \\
  18.668 $_{-0.003}^{+0.002}$ &     0.00 $_{}^{+    4.17}$ &    1.07 $\pm$ 0.31  &   56.8 $\pm$ 16.6  & 18.627 &    -40 $_{-  49}^{+  31}$ & 5.32$\times 10^{-03}$ &      O VII \\
  16.038 $_{-0.003}^{+0.008}$ &     8.89 $_{-    5.44}^{+    6.34}$ &    1.30 $_{-    0.81}^{+    0.59}$ &   48.0 $_{-   29.9}^{+   21.7}$ & 16.006 &    -100 $_{-  50}^{+ 150}$ & 3.84$\times 10^{-02}$ &     O VIII \\
  13.716 $_{-0.004}^{+0.003}$ &     8.34 $_{-    5.36}^{+    5.37}$ &    0.83 $_{-    0.20}^{+    0.23}$ &   17.9 $_{-    4.3}^{+    5.0}$ & 13.699 &   -330 $_{-  90}^{+  70}$ & 4.76$\times 10^{-05}$ &   Ne IX \\
  13.579 $_{-0.006}^{+0.005}$ &    13.48 $_{-    5.91}^{+    0.01}$ &    0.75 $_{-    0.19}^{+    0.18}$ &   16.3 $_{-    4.2}^{+    3.9}$ & 13.553 &   -117 $_{- 123}^{+ 116}$ & 1.15$\times 10^{-03}$ &  Ne IX \\
  12.155 $_{-0.004}^{+0.005}$ &     0.64 $_{-    0.64}^{+    8.48}$ &    0.41 $_{-    0.13}^{+    0.01}$ &   10.1 $_{-    3.2}^{+    0.2}$ & 12.134 &   -172 $_{-  92}^{+ 113}$ & 4.04$\times 10^{-03}$ &      Ne X \\
   8.467 $_{-0.002}^{+0.001}$ &     0.07 $_{-    0.07}^{+    3.10}$ &    0.21 $\pm$   0.05  &    5.7 $\pm$ 1.5  &--   &    -- & 1.80$\times 10^{-03}$ & (Mg XII ?)\\
   4.288 $_{-0.001}^{+0.004}$ &     0.00 $_{}^{+    5.87}$ &    0.23 $\pm$ 0.07  &    6.6 $\pm$  1.9  &  4.299 &  -1460 $_{-  40}^{+ 290}$ & 2.49$\times 10^{-03}$ &     S XV \\
   1.941 $_{-0.001}^{+0.002}$ &     1.48 $_{-    1.48}^{+    3.30}$ &    0.54 $_{-    0.12}^{+    0.17}$ &   10.8 $_{-    2.5}^{+    3.5}$ &  1.940 &   -620 $_{- 170}^{+ 230}$ & 4.51$\times 10^{-04}$ & Fe K$\alpha$ (narrow) \\
   1.927 $_{-0.012}^{+0.011}$ &    62.9 $_{-   11.1}^{+   14.4}$ &    2.31 $_{-    0.42}^{+    0.43}$ &   65.9 $_{-   11.9}^{+   12.2}$ &  1.940 &  -2720 $_{-1830}^{+1720}$ & 1.22$\times 10^{-06}$ &   Fe K$\alpha$ (broad) \\
   1.781 $\pm$ 0.004  &    10.00 $_{-    3.38}^{+    0.01}$ &    0.64 $_{-    0.17}^{+    0.16}$ &   17.9 $_{-    4.8}^{+    4.5}$ &  1.750 &   4530 $_{- 650}^{+ 700}$ & 5.23$\times 10^{-03}$ &     Fe K$\beta$ \\

 \enddata
\label{lineId}
\tablecomments{ The above table lists the properties of the emission lines measured in HETG coadded spectrum of NGC 4051. A range of ionization and velocity shifts are measured. The first column lists the initial wavelength. The second and third list the width as well as the normalization for the corresponding line. Those lines with a $\sigma$=0 are unresolved. The Equivalent width is given in the fourth column. $\lambda_0$ is the rest wavelength in NGC 4051 rest frame, which is at a z= 0.00234. The velocity is calculated using these rest wavelengths in column 6. The seventh column lists the significance of detection as 1 minus the confidence level. The final column lists particular ion that is being detected. All the line energies are fit to the data, which explains why the energy is different for the narrow and broad Fe K$\alpha$ lines.}
\end{deluxetable*}

%-------------------------------------Table End---------------------------------------------------------

%-------------------------------------Table Start---------------------------------------------------------
\begin{deluxetable*}{l l l l l l l l l}
\tablecolumns{9}
\tablewidth{0pc}
\tabletypesize{\scriptsize}
\tablecaption{Absorption Line Parameters}
\tablehead{ \colhead{line center} & \colhead{$\sigma $} & \colhead{Normalization} & \colhead{EW} & \colhead{$\lambda_\circ$ }& \colhead{velocity } & \colhead{Ftest} &\colhead{N$_z$ }& \colhead{ID} \\ (\AA\/) & (m\AA\/) & (10$^{-5}$ ph cm$^{-2}$ s$^{-1}$)  & (m\AA\/) & (\AA\/) & (km s$^{-1}$) & &(10$^{16}$ cm$^{-2}$) }
\startdata   
 21.617 $_{-0.008}^{+0.006}$ &    13.2 $_{-   13.2}^{+    9.6}$ &   -2.71 $_{-    0.66}^{+    0.52}$ &  -39.9 $_{-    9.7}^{+    7.6}$ & 21.602 &   -497 $_{- 111}^{+  79}$ & 3.04$\times 10^{-07}$ &      1.2 $\pm$  0.2  & O VII \\
  18.974 $_{-0.004}^{+0.001}$ &    15.7 $_{-    6.0}^{+    4.8}$ &   -3.46 $_{-    0.61}^{+    0.77}$ &  -48.2 $_{-    8.5}^{+   10.7}$ & 18.967 &   -584 $_{-  63}^{+  10}$ & 8.29$\times 10^{-35}$ &      4.8 $\pm$  0.8  & O VIII \\
  18.650 $_{-0.0012}^{}$ &    33.8 $_{-    5.2}^{+    5.7}$ &   -4.89 $_{-    0.64}^{+    0.61}$ &  -70.8 $_{-    9.3}^{+    8.9}$ & 18.627 &   -331 $_{-  19}^{}$ & 8.37$\times 10^{-28}$ &     14 $\pm$  2 & O VII \\
  17.790 $_{-0.008}^{+0.003}$ &     1.29 $_{-    1.29}^{+   12.86}$ &   -0.94 $_{-    0.58}^{+    0.17}$ &  -14.8 $_{-    9.1}^{+    2.6}$ & 17.768 &   -332 $_{- 126}^{+  43}$ & 6.11$\times 10^{-08}$ &      8.5 $_{-   1.3}^{+   1.4}$ & O VII \\
  16.020 $_{-0.0074}^{}$ &    14.3 $_{-    7.1}^{+    5.5}$ &   -1.98 $_{-    0.52}^{+    0.85}$ &  -34.5 $_{-    9.1}^{+   14.8}$ & 16.006 &   -438 $_{- 138}^{}$ & 3.94$\times 10^{-17}$ &     51 $\pm$  8 & O VIII \\
  15.266 $\pm$ 0.004 &     5.38 $_{-    5.38}^{+    4.79}$ &   -0.46 $_{-    0.15}^{+    0.14}$ &   -9.0 $_{-    3.0}^{+    2.8}$ & 15.261 &   -592 $_{-  76}^{+  77}$ & 4.17$\times 10^{-03}$ &      0.6 $\pm$ 0.1  & Fe XVII \\
  15.017 $\pm$ 0.002  &     4.98 $_{-    4.98}^{+    3.63}$ &   -0.76 $_{-    0.16}^{+    0.14}$ &  -15.3 $_{-    3.3}^{+    2.9}$ & 15.014 &   -641 $\pm$ 45  & 1.27$\times 10^{-11}$ &      0.3 $\pm$  0.1  & Fe XVII \\
  14.220 $\pm$ 0.006  &    28.2 $_{-    5.8}^{+    6.7}$ &   -1.55 $_{-    0.32}^{+    0.29}$ &  -32.7 $_{-    6.7}^{+    6.2}$ & 14.203 &   -334 $_{- 137}^{+ 132}$ & 2.06$\times 10^{-08}$ &      1.7 $\pm$ 0.3 & Fe XVIII \\
  14.042 $_{-0.013}^{+0.010}$ &    23.4 $_{-    6.1}^{+   10.4}$ &   -0.83 $_{-    0.29}^{+    0.25}$ &  -17.9 $_{-    6.3}^{+    5.3}$ & -- & -- & 4.79$\times 10^{-03}$ &    -- & Unidentified \\
  13.841 $\pm$ 0.002   &     8.41 $_{-    3.57}^{+    4.51}$ &   -0.76 $_{-    0.18}^{+    0.15}$ &  -16.1 $_{-    3.8}^{+    3.1}$ & 13.844 &   -751 $_{-  49}^{+  52}$ & 1.53$\times 10^{-09}$ &     16.2 $\pm$ 2.6  & Fe XIX \\
  13.658 $_{-0.005}^{+0.003}$ &    14.3 $_{-    5.3}^{+    8.3}$ &   -0.56 $_{-    0.19}^{+    0.17}$ &  -12.1 $_{-    4.2}^{+    3.7}$ & 13.670 &   -971 $_{- 119}^{+  58}$ & 4.47$\times 10^{-03}$ &      1.4 $\pm$  0.2  & Fe XIX \\
  13.515 $\pm$ 0.002  &     8.18 $_{-    2.91}^{+    3.19}$ &   -0.73 $_{-    0.14}^{+    0.13}$ &  -16.0 $_{-    3.1}^{+    2.8}$ & 13.518 &   -769 $_{-  51}^{+  53}$ & 8.30$\times 10^{-11}$ &      1.1 $\pm$   0.2 & Fe XIX \\
  13.450 $\pm$ 0.002  &    14.6 $_{-    2.5}^{+    2.1}$ &   -1.23 $_{-    0.16}^{+    0.15}$ &  -26.9 $_{-    3.5}^{+    3.3}$ & 13.447 &   -629 $_{-  52}^{+  51}$ & 6.48$\times 10^{-22}$ &      2.1 $\pm$  0.3  & Ne IX \\
  12.832 $\pm$ 0.003 &     9.78 $_{-    2.99}^{+    3.34}$ &   -0.66 $_{-    0.12}^{+    0.11}$ &  -15.1 $_{-    2.8}^{+    2.6}$ & 12.847 &  -1046 $_{-  62}^{+  67}$ & 2.33$\times 10^{-10}$ &      2.2 $\pm$  0.4  & Fe XX \\
  12.281 $_{-0.004}^{+0.003}$ &     3.91 $_{-    3.91}^{+    9.75}$ &   -0.35 $_{-    0.18}^{+    0.09}$ &   -8.3 $_{-    4.3}^{+    2.1}$ & 12.285 &   -779 $_{- 106}^{+  65}$ & 1.07$\times 10^{-04}$ &      0.6 $\pm$  0.1  & Fe XXI \\
  12.132 $_{-0.002}^{+0.012}$ &     9.12 $_{-    2.72}^{+    6.05}$ &   -0.87 $_{-    0.48}^{+    0.14}$ &  -20.9 $_{-   11.4}^{+    3.4}$ & 12.134 &   -737 $_{-  49}^{+ 303}$ & 7.62$\times 10^{-25}$ &      5.1 $\pm$  0.8  & Ne X \\
 11.935 $_{-0.0002}^{+0.045}$ &     0.00 $_{}^{+    6.86}$ &   -0.13 $\pm$ 0.07  &   -3.1 $\pm$ 1.7  & 11.920 &   -320 $_{-   4}^{+1100}$ & 4.6$\times 10^{-02}$ &      0.4 $\pm$ 0.1   & Fe XXII \\
  11.771 $\pm$ 0.002  &     2.90 $_{-    2.90}^{+    4.29}$ &   -0.32 $_{-    0.08}^{+    0.07}$ &   -8.1 $_{-    2.1}^{+    1.7}$ & 11.770 &   -687 $_{-  56}^{+  55}$ & 3.54$\times 10^{-05}$ &      0.9 $\pm$  0.1  & Fe XXII \\
  11.655 $_{-0.005}^{+0.001}$ &     0.16 $_{-    0.16}^{+    4.44}$ &   -0.19 $\pm$ 0.06  &   -4.7 $\pm$  1.6  & -- & -- & 4.40$\times 10^{-02}$ &   -- & Unidentified\\
  11.551 $\pm$ 0.003  &     6.11 $_{-    6.11}^{+    3.55}$ &   -0.33 $\pm$   0.08 &   -8.4 $\pm$  2.1 & 11.547 &   -602 $\pm$  74 & 1.01$\times 10^{-04}$ &      4.2 $\pm$ 0.7 & Ne IX \\
  11.415 $_{-0.0001}^{+0.005}$ &     0.00 $_{}^{+    4.50}$ &   -0.18 $\pm$  0.06 &   -4.6 $\pm$  1.5 & 11.427 &  -1014 $_{-   1}^{+ 131}$ & 3.48$\times 10^{-02}$ &      0.2 $\pm$ 0.1 & Fe XXII \\
  11.001 $\pm$ 0.004 &     8.03 $_{-    4.35}^{+    4.15}$ &   -0.25 $\pm$  0.08 &   -6.3 $_{-    2.0}^{+    1.9}$ & 11.001 &   -691 $_{- 121}^{+ 120}$ & 1.13$\times 10^{-02}$ &      9.3 $\pm$ 1.5 & Ne IX \\
  10.237 $_{-0.004}^{+0.003}$ &     5.41 $_{-    5.41}^{+    4.84}$ &   -0.25 $_{-    0.08}^{+    0.07}$ &   -6.4 $_{-    1.9}^{+    1.6}$ & 10.238 &   -738 $_{- 103}^{+  85}$ & 5.60$\times 10^{-04}$ &     11.5 $\pm$ 1.8 & Ne X \\
   9.838 $_{-0.0003}^{+0.002}$ &     0.00 $_{}^{+    3.03}$ &   -0.18 $\pm$ 0.05 &   -4.8 $\pm$ 1.3 &  --&  --& 3.53$\times 10^{-03 }$&   --  & (Fe XXI?)\\
   9.174 $_{-0.004}^{+0.003}$ &     8.19 $_{-    3.33}^{+    4.65}$ &   -0.28 $_{-    0.01}^{+    0.08}$ &   -7.7 $_{-    0.3}^{+    2.3}$ &  9.169 &   -529 $_{- 118}^{+ 110}$ & 6.97$\times 10^{-04}$ &      1.2 $\pm$ 0.2 & Mg XI \\
   8.421 $\pm$ 0.001 &     2.44 $_{-    2.44}^{+    1.84}$ &   -0.26 $\pm$ 0.05 &   -7.2 $\pm$ 1.4&  8.419 &   -635 $_{-  34}^{+  36}$ & 2.23$\times 10^{-13}$ &      3.7 $\pm$ 0.6 & Mg XII \\
   6.650 $_{-0.0018}^{+0.0024}$ &     4.32 $_{-    2.60}^{+    2.84}$ &   -0.23 $_{-    0.07}^{+    0.06}$ &   -6.5 $_{-    2.0}^{+    1.6}$ &  6.648 &   -612 $_{-  79}^{+ 109}$ & 3.34$\times 10^{-05}$ &      1.9 $\pm$  0.3 & Si XIII \\

      \enddata
\label{lineIdabs}
\tablecomments{ This table lists the absorption lines detected in our phenomenological modeling in the HETG spectrum of NGC 4051. All of the features appear to be blue shifted with respect to the host galaxy, indicating a bulk outflow from the galaxy. The first column lists the initial wavelength. The second and third list the width as well as the normalization for the corresponding line. A negative normalization indicates an absorption feature. Those lines with a $\sigma$=0 are unresolved. The Equivalent width is given in the fourth column. $\lambda_0$ is the rest wavelength in NGC 4051 rest frame, which is at a z= 0.00234. The velocity is calculated using these rest wavelengths in column 6. The seventh column lists the significance of detection as 1 minus the confidence level. The eighth column lists the column densities derived using the relation $W_{\lambda}=(\pi e^2 / m_ec^2) N_j \lambda^2 f_{ij}$, where $W_\lambda$ is the equivalent width, $N_j$ is the column density, $\lambda$ is the rest wavelength, and $f_{ij}$ is the oscillator strength. The final column lists particular ion that is being detected. All the line energies are fit to the data.}

\end{deluxetable*}

%-------------------------------------Table End---------------------------------------------------------

%-------------------------------------Table Start---------------------------------------------------------
\begin{deluxetable*}{c c c | c  | c c | c c}
\tablecolumns{8}
\tablewidth{0pt}
\tablecaption{XSTAR Model}
\tablehead{ & & & Absorption & \multicolumn{2}{|c|}{Emission} && \\ \hline  \colhead{Density} & \colhead{Column Density }  & \colhead{$\log (\xi) $} &\colhead{velocity} & \colhead{Normalization} & \colhead{velocity} & \colhead{$\Delta \chi^2$}  &\colhead{ Ftest}  \\ (cm$^{-3}$) &  (10$^{20}$ cm$^{-2}$) & (ergs cm s$^{-1}$) &  (km s$^{-1}$)  & (10$^{-3}$) & (km s$^{-1}$) &&  }
\startdata
10$^{11}$&  8.10$^{+  0.92}_{ -0.99}$&   4.50$^{}_{ -0.06}$&   -680$^{+    20}_{   -60}$&  19.53$^{+  5.22}_{ -5.08}$&    -50$^{+   150}_{  -250}$& 106 & 6.3$\times 10^{-21}$\\
10$^{11}$& 10.10$^{+  0.63}_{ -0.95}$&   3.28$^{+  0.05}_{ -0.03}$&   -630$^{+    10}_{   -80}$&   1.01$^{+  0.59}_{ -0.44}$&    100$^{+   100}_{  -110}$& 1076 & 1.8$\times 10^{-193}$\\
10$^{10}$&  1.97$^{+  0.27}_{ -0.25}$&   1.00$^{+  0.06}_{ -0.15}$&   -400$^{+   380}_{  -270}$&   1.08$^{+  0.27}_{ -0.16}$&   -300$^{+    70}_{   -40}$ & 248 & 26.5$\times 10^{-48}$\\

\hline \hline
Power-law & Normalization & Constant & Diskbb & Normalization  & Phabs \\ 
($\Gamma$) & (10$^{-3}$ ph &  & (keV)  & $\tiny{\left( \left( \frac{R_{in}/km}{D/kpc} \right)^2\cos \theta \right)}$ &$\times10^{20}$ \\
&cm$^{-2}$ s$^{-1}$ keV$^{-1}$) & &&& (atoms cm$^{-2}$) \\
 \hline
  1.95 $\pm$ 0.01&   4.32$^{+  0.04}_{ -0.03}$&  1.059$^{+ 0.005}_{-0.004}$ &  0.133$^{+ 0.002}_{-0.003}$&  3.78$^{+ 0.40}_{-0.37}\times10^{3}$& 1.15\\

\enddata  

\label{xstar}
\tablecomments{ The above table lists the parameters of the three XSTAR model components. The errors reported here are $1 \sigma$. $\chi^2 / \nu$ 8177/7516=1.09. The actual model is as follows: phabs*(mtable1*mtable2*mtable3(power-law+diskbb)+atable1 + atable2 + atable3)*constant) The multiplicative constant is applied to the MEG spectrum to correct for the slight discrepancy between the flux in the two spectrographs. * Denotes the model reached a maximum in the ionization parameter limits}
\end{deluxetable*}
%-------------------------------------Table End---------------------------------------------------------

%-------------------------------------Table Start---------------------------------------------------------
\begin{deluxetable*}{c c c |  c | c c | c c}
\tablecolumns{8}
\tablewidth{0pc}
\tablecaption{Cloudy Model}
\tablehead{& & & Absorption & \multicolumn{2}{|c|}{Emission} && \\ \hline  \colhead{Density} & \colhead{Column Density }  & \colhead{Radius} &\colhead{velocity} & \colhead{Normalization} & \colhead{velocity} & \colhead{$\Delta \chi^2$}  &\colhead{ Ftest}  \\ (cm$^{-3}$) &  (10$^{20}$ cm$^{-2}$) & (10$^{13}$ cm) &  (km s$^{-1}$)  & (10$^{-21}$) & (km s$^{-1}$) &&  }
\startdata
10$^{11}$& 21.8$^{+  4.4}_{ -3.5}$&   1.00$^{+  0.04}_{}$&  -1090$^{+   110}_{  -120}$ &  0.007$\pm$ 0.002 &   -110$^{+    80}_{  -130}$& 23 & 1.4 $\times$ 10$^{-3}$ \\
10$^{11}$&  7.68$^{+  0.84}_{ -0.75}$&   5.09$^{+  0.29}_{ -0.26}$ &  -1000$^{+    70}_{   -30}$ &  0.015$\pm$ 0.006 &    220$^{+   180}_{   -60}$& 642 & 7.6 $\times$ 10$^{-96}$ \\
10$^{10}$&  6.42$^{+  0.65}_{ -0.81}$&  82.9$^{+  4.7}_{ -4.4}$ &   -700$^{+    30}_{   -20}$ &  2.47$^{+ 0.80}_{ -0.70}$&   -290$^{+    60}_{  -180}$& 484 & 1.9 $\times$ 10$^{-54}$\\

\hline \hline
Power-law & Normalization & Constant & Diskbb & Normalization  & Phabs \\ 
($\Gamma$) & (10$^{-3}$ photons  &  & (keV)  & $\tiny{\left( \left( \frac{R_{in}/km}{D/kpc} \right)^2\cos \theta \right)}$ &$\times10^{20}$ \\
&cm$^{-2}$ s$^{-1}$ keV$^{-1}$) & &&& (atoms cm$^{-2}$) \\
 \hline
   1.97 $\pm$ 0.01 &   4.42 $^{+  0.04}_{ -0.03}$&  1.06 $\pm$ 0.01&  0.127 $^{+ 0.002}_{-0.003}$&  4.81 $^{+ 0.77}_{-0.57}\times10^{3}$ & 1.15  \\
\enddata  
\label{cloudy}
\tablecomments{ This table lists the Cloudy model parameters of the three components required by the data. The errors reported here are $1 \sigma$. $\chi^2 / \nu $ =8713/7516=1.16. The actual model is as follows: phabs*(mtable1*mtable2*mtable3(power-law+diskbb)+atable1 + atable2 + atable3)*constant). The multiplicative constant is applied to the MEG spectrum to correct for the slight discrepancy between the flux in the two spectrographs. }
\end{deluxetable*}

%-------------------------------------Table End---------------------------------------------------------

%-------------------------------------Table Start---------------------------------------------------------
\begin{deluxetable*}{l | l l | l l | l l}
\tablecolumns{7}
\tablewidth{0pc}
\tablecaption{Photoionization Model}
\tablehead{ & XSTAR & Cloudy & XSTAR & Cloudy & XSTAR & Cloudy }
\startdata
Density (cm$^{-3}$) & \multicolumn{2}{c}{$10^{11}$} & \multicolumn{2}{c}{$10^{11}$} & \multicolumn{2}{c}{$10^{10}$}\\ 
$\log \xi$ (ergs cm$^{-1}$) &  4.50 $^{*}_{ -0.06}$&   5.00 $^{*}_{ -0.03}$&   3.28 $^{+  0.05}_{ -0.03}$&   3.59 $\pm$ 0.05 &   1.00 $^{+  0.06}_{ -0.15}$&   2.16 $\pm$ 0.05 \\
Radius ($10^{13}$cm) &  1.78 $^{+  0.12}_{*}$&   1.00 $^{+  0.04}_{ *}$&   7.25 $^{+  0.29}_{ -0.38}$&   5.09 $^{+  0.29}_{ -0.26}$& 317 $^{+ 58}_{-20}$&  82.9 $^{+  4.7}_{ -4.4}$\\
Radius (R$_g$) &      71 $^{+       4}_{  *}$&       40 $^{+       1}_{       *}$&      290 $^{+      11}_{     -15}$&      203 $^{+      11}_{     -10}$&    12700 $^{+    2300}_{    -800}$&     3310 $^{+     190}_{    -170}$\\
$\Delta$(r) (10$^{10}$cm) &  0.81 $^{+  0.09}_{ -0.10}$&   2.18 $^{+  0.44}_{ -0.35}$&   1.01 $^{+  0.06}_{ -0.09}$&   0.77 $\pm$ 0.08 &   0.20 $\pm$ 0.03 &   0.64 $^{+  0.07}_{ -0.08}$\\
$C_v$ (10$^{-4}$)&      4.56 $^{*}_{      0.81}$&      21.8 $^{*}_{      4.2}$&       1.39 $^{+      0.17}_{      0.18}$&       1.51 $^{+      0.25}_{      0.22}$&       0.06 $^{+      0.01}_{      0.02}$&       0.77 $\pm$  0.13 \\
$\dot{M}_{out}$ ($10^{-6}$M$_{\odot}$yr$^{-1}$ )&      0.30 $^{+      0.09}_{  *}$&       0.73 $^{+      0.28}_{   *}$&       1.42 $^{+      0.35}_{     -0.22}$&       1.20 $^{+      3.39}_{ -     1.82}$&       7.7 $^{+      9.6}_{     -7.4}$&      11.5 $^{+   7.1}_{     -8.6}$\\
 \enddata  
\label{photo}
\tablecomments{ This table compares the XSTAR and Cloudy model components. The models give similar values, and consistent mass outflow rates. We note that in general the lowest ionization component carries the most mass. We have assumed 1.7 $\pm$ $0.5 \times 10^6 M_\odot$ as the mass of the SMBH. The symbol, *, denotes both upper and lower limits in the models.}
\end{deluxetable*}

%-------------------------------------Table End---------------------------------------------------------

%-------------------------------------Table Start---------------------------------------------------------
\begin{deluxetable*}{c | c c c c c c c}
%\rotate
\tablecolumns{8}
\tablecaption{Comparing Models}
\tablehead{ &\scriptsize{Our XSTAR} & \scriptsize{Collinge}& \scriptsize{Krongold} &\scriptsize{Steenbrugge}  &\scriptsize{Nucita} & \scriptsize{Lobban} & \scriptsize{Pounds}  \\ & \scriptsize{Model} & \scriptsize{et al. (2001)} & \scriptsize{et al. (2007)}  & \scriptsize{et al. (2009)}  & \scriptsize{et al. (2010)}  & \scriptsize{et al. (2011)}  & \scriptsize{et al. (2011)} \\ Instrument & (HETG) & (HETG) & (RGS )  & (LETGS) ** & (RGS) & (HETG) & (RGS) }
\startdata
$\log \xi_0$ (ergs cm s$^{-1}$) & -  & -&  - & 0.07 $\pm$ 0.13 & - & -0.86 $\pm$ 0.30 & 0.32 $\pm$ 24 \\
v$_0$ (km s$^{-1}$) & &&& -210 $\pm$ 70 & &-180 $\pm$ 100  & +120 $\pm$ 60\\
$\log n_0$ (cm$^{-3}$)&&& &$<$ 6.5  && - & -\\
$\log r_0$ (cm) && &&  $>$ 17.95 &  &$\lesssim$ 23.4 & -\\
$\log$ N$_{H,0}$ (cm$^{-2}$)& &&&20.08 $\pm$ 0.18 && 20.48 $\pm$ 0.13 & 20.0 $\pm$ 0.1\\
\\
$\log \xi_1$(ergs cm s$^{-1}$) &1.00 $^{+0.06}_{-0.15}$ & - & - & 0.87 $\pm$ 0.09 & - & 0.60 $\pm$ 0.30 & 1.43 $\pm$ 0.22 \\
v$_1$ (km s$^{-1}$) &-400 $^{+380}_{-270}$ & -600 $\pm$ 130 & &-200 $\pm$ 30 & & -220 $\pm$ 60  & -530 $\pm$ 90\\
$\log n_1$ (cm$^{-3}$)  & 10& $<$ 10.6 && $>$ 6.6 & & $<$ 11 & $\sim$ 6.7 \\
$\log r_1$ (cm) &15.5 $\pm$ 0.1& - && $<$ 17.48 & &15.8 -- 22.6 &$\sim$ 18\\
$\log$ N$_{H,1}$ (cm$^{-1}$) & 20.3 $\pm$ 0.1& $>$ 21.3   && 20.46 $\pm$ 0.08 & &20.31 $\pm$ 0.13 &  20.04 $\pm$ 0.26\\
\\
$\log \xi_{2a}$(ergs cm s$^{-1}$) & - & - & 2.18 $\pm$ 0.02 * & 2.32 $\pm$ 0.13 & 2.15 $\pm$ 0.02 & 1.96 $\pm$ 0.17 &  2.52 $\pm$ 0.07 \\
v$_{2a}$ (km s$^{-1}$) & &&-492 $\pm$ 97 &-580 $\pm$ 50 & 0 &-820 $\pm$ 30 & -3850 $\pm$ 60 \\
$\log n_{2a}$ (cm$^{-3}$) & & & $>$ 7.9 &$<$ 6.7&$ \lesssim$ 7 &- & $\sim$ 5.7\\
$\log r_{2a}$ (cm) & && $<$ 14.9& $>$ 16.70 & $\gtrsim$ 16.8 & $\lesssim$ 20.3 & $\sim$ 17.8 \\
$\log$ N$_{H,2a}$ (cm$^{-2}$) && & 20.73 $\pm$ 0.17 & 20.90 $\pm$ 0.3 & 23.36 $\pm$ 0.01 &20.70 $\pm$ 0.16 & 21.26 $\pm$ 0.10\\
\\
$\log \xi_{2b}$(ergs cm s$^{-1}$) &- & - & - & - &  - &2.16 $\pm$ 0.21 & 2.77 $\pm$ 0.15\\
v$_{2b}$ (km s$^{-1}$) & && & &&-550 $\pm$ 60 & -400 $\pm$ 50 \\
$\log n_{2b}$ (cm$^{-3}$) &  &&&&& - & -\\
$\log r_{2b}$ (cm) & &&& &&$\lesssim$ 20.4 & -   \\
$\log$ N$_{H,2b}$ (cm$^{-2}$)& & & &&& 21.04 $\pm$ 0.27 & 21.46 $\pm$ 0.29\\
\\
$\log \xi_3$(ergs cm s$^{-1}$) &3.28 $^{+0.05}_{0.03}$&- &- &  3.19 $\pm$ 0.09 & - &2.97 $\pm$ 0.10 & 2.97 $\pm$ 0.06\\
v$_3$ (km s$^{-1}$) &-620 $^{+10}_{-80}$ & -2340 $\pm$ 130 & &-4670 $\pm$ 150 & &-710 $\pm$ 40 & -5880 $\pm$ 60\\
$\log n_3$ (cm$^{-3}$) & 11 & $<$ 10.6 & &$<$ 5.4  & &- &  $\sim$ 5.7  \\
$\log r_3$ (cm)  & 13.86 $\pm$ 0.02 & - && $>$ 16.85 & & $\lesssim$ 18.7 & $\sim$ 17.8 \\
$\log$ N$_{H,3}$ (cm$^{-2}$) &21.00 $^{+0.03}_{-0.04}$ & 19 -- 20.7 && 22.30 $\pm$ 0.3 & & 21.44 $\pm$ 0.12 & 22.15 $\pm$ 0.07\\
\\
$\log \xi_4$(ergs cm s$^{-1}$) & $\gtrsim$4.5 & - &4.42 $\pm$ 0.07 * &-  &- & 4.1 $\pm$ 0.2 & -\\
v$_4$ (km s$^{-1}$)  & -680 $^{+20}_{-60}$ && -537 $\pm$ 130 & & &-5800 $\pm$ 1200 \\
$\log n_4$ (cm$^{-3}$) & 11 && 6.8 -- 7.3 & & &6.8 -- 8.8   \\
$\log r_4$ (cm) & 13.25 $^{+0.03}$ && 15.1 -- 15.4 & && 15.1 -- 16.1\\
$\log$ N$_{H,4}$ (cm$^{-2}$)& 20.91 $^{+0.05}_{-0.06}$ & &21.42 $\pm$ 0.12 & & &22.92 $\pm$ 0.12\\
\\  

 \enddata  
\label{compare}
\tablecomments{This table shows the comparison between other absorption X-ray components found in the literature and the work presented in this paper. *\cite{Krongold07} does not use $\xi$ but instead uses the ionization parameter of U = $\frac{Q(H)}{4\pi r^2 n_e c}$, where Q(H) is the number of ionizing photons, r is the distance from the ionizing source, $n_e$ is the electron number density and c is the speed of light. To convert between the two, we assumed an ionizing luminosity of 10$^{42}$. ** Used Spectrum C for this comparison. }
\end{deluxetable*}
%-------------------------------------Table End---------------------------------------------------------

\clearpage
%------------------------------------Figure Start---------------------------------------------------------
%------------------------------------------------------------------------------------------------------------

%------------------------------------Figure Start---------------------------------------------------------
%------------------------------------------------------------------------------------------------------------

%------------------------------------Figure Start---------------------------------------------------------
%------------------------------------------------------------------------------------------------------------

%------------------------------------Figure Start---------------------------------------------------------
%------------------------------------------------------------------------------------------------------------

%------------------------------------Figure Start---------------------------------------------------------
\begin{figure}
\begin{center}
\includegraphics[height=3in,angle=270]{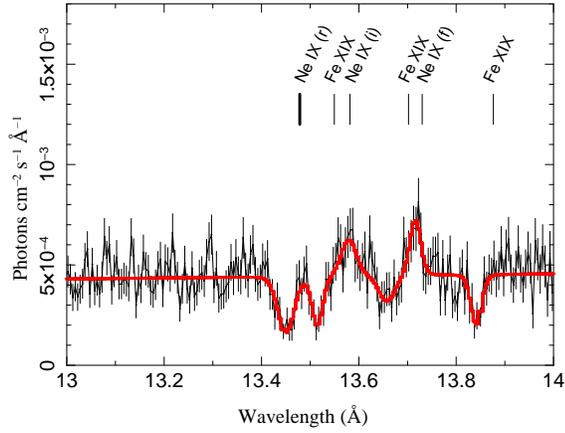}
\caption{\footnotesize{ This figure shows the Ne IX line complex that is modeled with Gaussian lines for both the absorption and emission. This He-like triplet gives a density constraint of $n_e<$10$^{12}$ cm$^{-3}$.  Line identifications are in the rest frame of NGC 4051 (z= 0.002336).The wavelength plotted is in the observed reference frame. }}
\label{NeIX}
\end{center}
\end{figure}
%------------------------------------------------------------------------------------------------------------

%------------------------------------Figure Start---------------------------------------------------------
\begin{figure}
\begin{center}
\includegraphics[height=3in,angle=270]{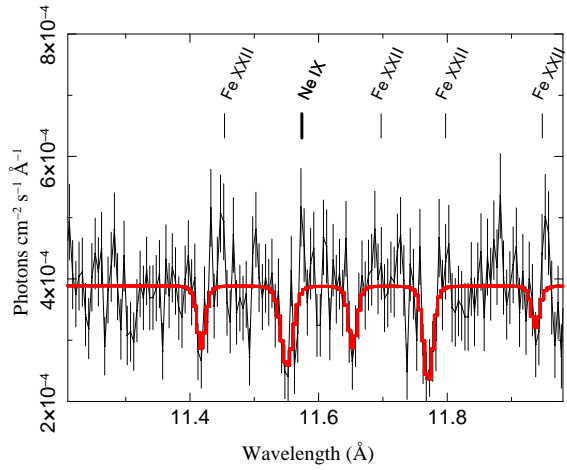}
\caption{\footnotesize{This figure shows the Fe XXII lines, which are modeled with Gaussian profiles. These lines were used as density diagnostics, constraining the density to be $n_e \lesssim$ 7 $\times$ 10$^{12}$ cm$^{-3}$.The wavelength plotted is in the observed reference frame, while the line identifications are in the rest frame of the galaxy. }}
\label{FeXXII}
\end{center}
\end{figure}
%------------------------------------------------------------------------------------------------------------

%------------------------------------Figure Start---------------------------------------------------------
\begin{figure}
\begin{center}
\includegraphics[width=3in]{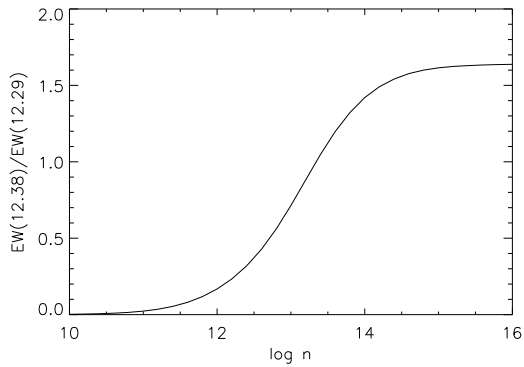}
\caption{\footnotesize{This figure shows how the ratio of Fe XXI lines 12.38/12.285 change with increasing density. This plot was created using the CHIANTI populations at a temperature of 5 $\times 10^5$ K.}}
\label{chiantife}
\end{center}
\end{figure}
%------------------------------------------------------------------------------------------------------------

%------------------------------------Figure Start---------------------------------------------------------
\begin{figure}
\begin{center}
\includegraphics[height=3in,angle=270]{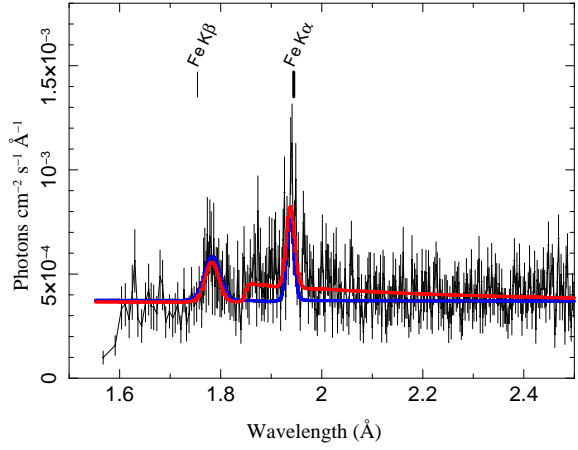}
\caption{\footnotesize{The above plot shows the fluorescence Fe K$\alpha$ line in NGC 4051, as well as two narrow Fe K $\alpha$ and $\beta$ lines. The narrow lines likely originate in distant material, while the broad fluorescence line likely originates in the inner-most parts of the accretion disk. The red line denotes the relativistic fluorescence line characterized with the \texttt{Kerrdisk} line profile. The blue line does not include this line. The narrow Fe K $\alpha$ and $\beta$ features are modeled with Gaussian lines. The line markers are drawn in the rest frame of NGC 4051. The wavelength plotted is in the observed reference frame. }}
\label{fefl}
\end{center}
\end{figure}
%------------------------------------------------------------------------------------------------------------

%------------------------------------Figure Start---------------------------------------------------------
\begin{figure}
\begin{center}
\includegraphics[height=3in,angle=270]{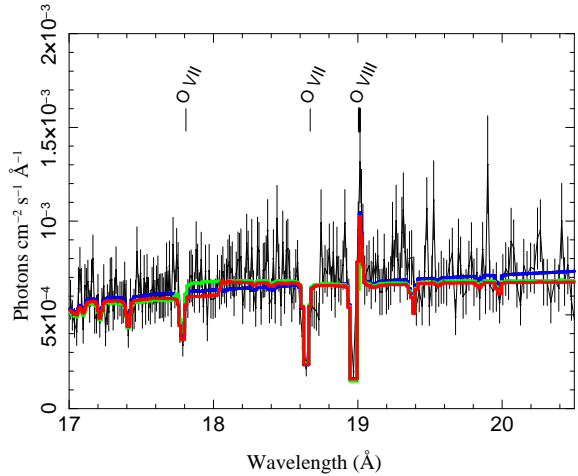}
\caption{\footnotesize{This is the O VIII Lyman $\alpha$ fluorescence line modeled with a  \texttt{Kerrdisk} component to describe any relativistic O VIII emission. The blue line is model with no relativistic line but includes the XSTAR model. The green line includes a relativistic line with all of its parameters to vary ( spin, inclination angle, inner radius, and normalization). The red line denotes the model in which the O VIII parameters were tied to that of the Fe K$\alpha$ line. The line is suggested at a 2 $\sigma$ level if the parameters are free to vary or tied to that of the Fe K$\alpha$ line. The wavelength plotted is in the observed reference frame, and the data were rebinned with a minimum of 25 counts per bin for clarity. }}
\label{okerr}
\end{center}
\end{figure}
%------------------------------------------------------------------------------------------------------------

%------------------------------------Figure Start---------------------------------------------------------
\begin{figure*}[t]
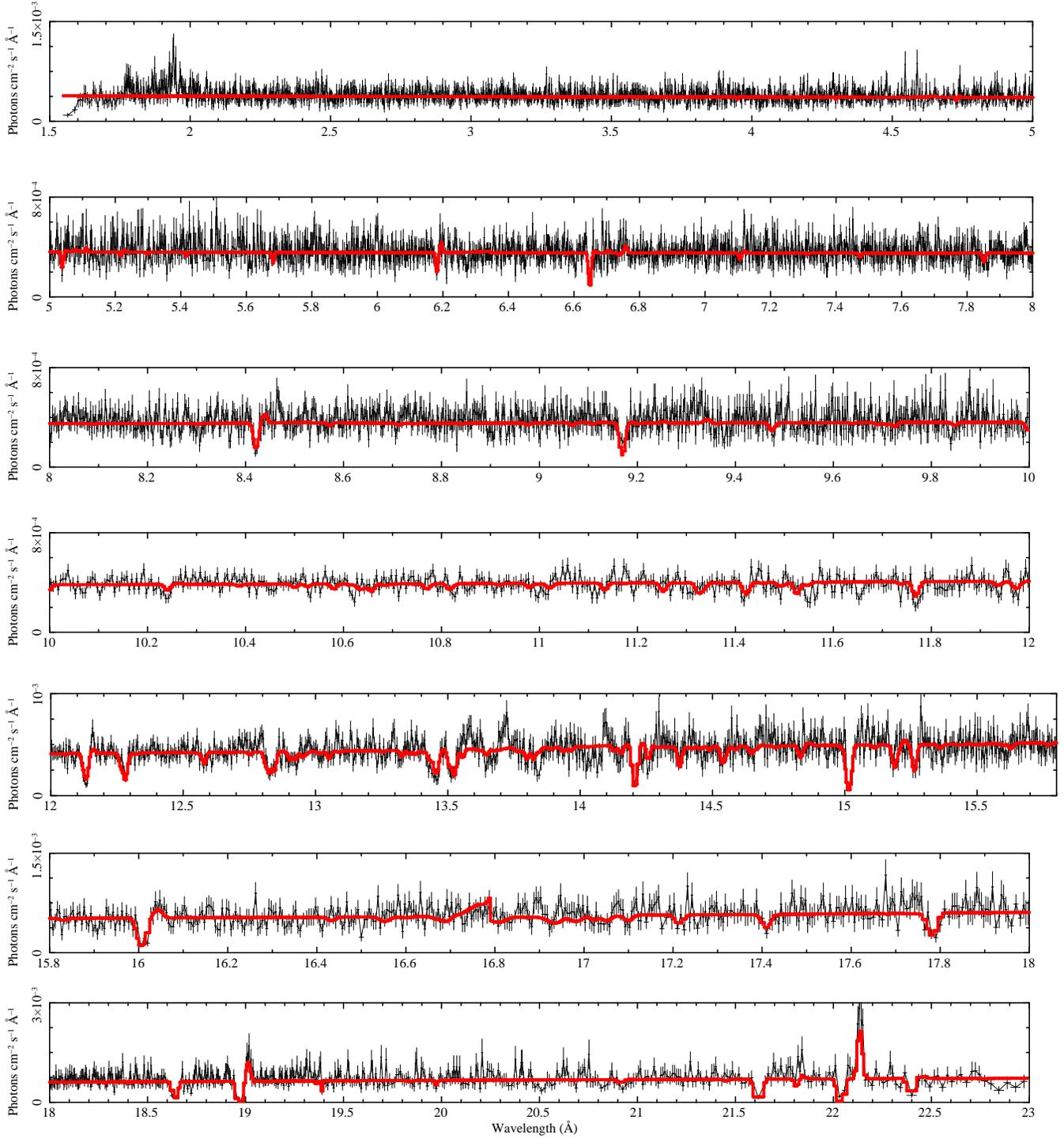

\begin{center}
\includegraphics[height=7in,angle=270]{f16.ps}
\includegraphics[height=7in,angle=270]{f17.ps}
\includegraphics[height=7in,angle=270]{f18.ps}
\includegraphics[height=7in,angle=270]{f19.ps}
\includegraphics[height=7in,angle=270]{f20.ps}
\includegraphics[height=7in,angle=270]{f21.ps}
\includegraphics[height=7in,angle=270]{f22.ps}
\caption{\footnotesize{This figure shows the HETG spectra and the best-fit XSTAR model components.  The parameters for this model are given in Table \ref{xstar}, and the model has $\chi^2 / \nu$ = 8177/7516. The absorption features in the residuals as given in Figure \ref{ratiorange} are well characterized by the XSTAR model, especially in the ranges of 12--18\AA\/. The wavelength plotted is in the observed reference frame. }}
\label{xstarrange}
\end{center}
\end{figure*}
%------------------------------------------------------------------------------------------------------------
%------------------------------------Figure Start---------------------------------------------------------
\begin{figure*}[t]
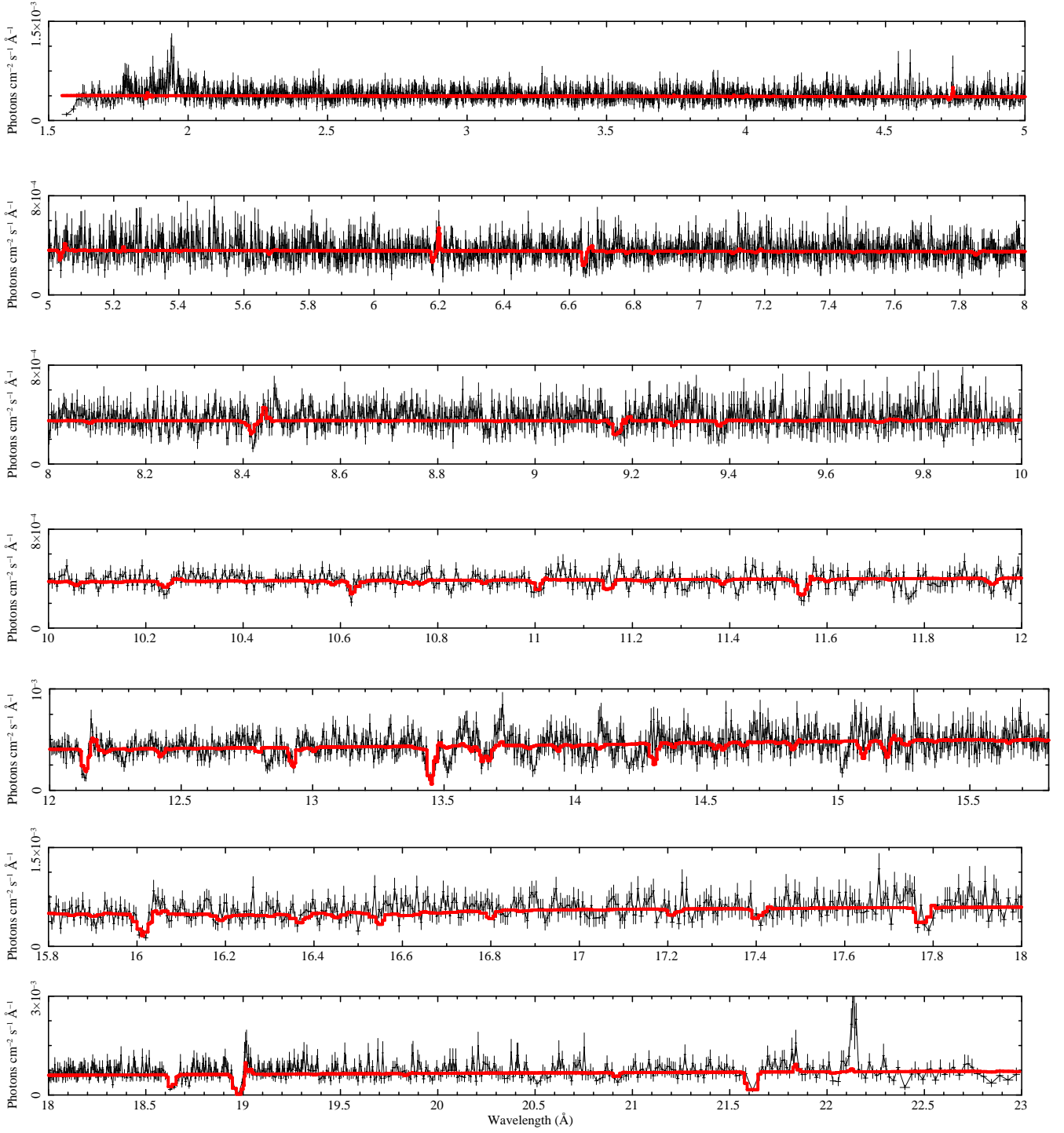

\begin{center}
\includegraphics[height=7in,angle=270]{f23.ps}
\includegraphics[height=7in,angle=270]{f24.ps}
\includegraphics[height=7in,angle=270]{f25.ps}
\includegraphics[height=7in,angle=270]{f26.ps}
\includegraphics[height=7in,angle=270]{f27.ps}
\includegraphics[height=7in,angle=270]{f28.ps}
\includegraphics[height=7in,angle=270]{f29.ps}
\caption{\footnotesize{The above figure plots the HETG spectra and the best-fit Cloudy model. The parameters for this model are given in Table \ref{cloudy} and the model has $\chi^2 / \nu$ = 8713/7516. The wavelength plotted is in the observed reference frame. }}
\label{cloudyrange}
\end{center}
\end{figure*}
%------------------------------------------------------------------------------------------------------------

%\bibliography{bib4051}

\begin{thebibliography}{48}
\expandafter\ifx\csname natexlab\endcsname\relax\def\natexlab#1{#1}\fi

\bibitem[{{Antonucci}(1993)}]{Antonucci93}
{Antonucci}, R. 1993, \araa, 31, 473

\bibitem[{{Behar}(2009)}]{Behar09}
{Behar}, E. 2009, \apj, 703, 1346

\bibitem[{{Blandford} \& {Payne}(1982)}]{Blandford82}
{Blandford}, R.~D., \& {Payne}, D.~G. 1982, \mnras, 199, 883

\bibitem[{{Blum} {et~al.}(2010){Blum}, {Miller}, {Cackett}, {Yamaoka},
  {Takahashi}, {Raymond}, {Reynolds}, \& {Fabian}}]{Blum10}
{Blum}, J.~L., {Miller}, J.~M., {Cackett}, E., {Yamaoka}, K., {Takahashi}, H.,
  {Raymond}, J., {Reynolds}, C.~S., \& {Fabian}, A.~C. 2010, \apj, 713, 1244

\bibitem[{{Blustin} {et~al.}(2005){Blustin}, {Page}, {Fuerst},
  {Branduardi-Raymont}, \& {Ashton}}]{Blustin05}
{Blustin}, A.~J., {Page}, M.~J., {Fuerst}, S.~V., {Branduardi-Raymont}, G., \&
  {Ashton}, C.~E. 2005, \aap, 431, 111

\bibitem[{{Brenneman} \& {Reynolds}(2006)}]{Brenneman06}
{Brenneman}, L.~W., \& {Reynolds}, C.~S. 2006, \apj, 652, 1028

\bibitem[{{Chartas} {et~al.}(2009){Chartas}, {Kochanek}, {Dai}, {Poindexter},
  \& {Garmire}}]{Chartas09}
{Chartas}, G., {Kochanek}, C.~S., {Dai}, X., {Poindexter}, S., \& {Garmire}, G.
  2009, \apj, 693, 174

\bibitem[{{Collinge} {et~al.}(2001){Collinge}, {Brandt}, {Kaspi}, {Crenshaw},
  {Elvis}, {Kraemer}, {Reynolds}, {Sambruna}, \& {Wills}}]{Collinge01}
{Collinge}, M.~J., {et~al.} 2001, \apj, 557, 2

\bibitem[{{Crummy} {et~al.}(2006){Crummy}, {Fabian}, {Gallo}, \&
  {Ross}}]{Crummy06}
{Crummy}, J., {Fabian}, A.~C., {Gallo}, L., \& {Ross}, R.~R. 2006, \mnras, 365,
  1067

\bibitem[{{Denney} {et~al.}(2009){Denney}, {Watson}, {Peterson}, {Pogge},
  {Atlee}, {Bentz}, {Bird}, {Brokofsky}, {Comins}, {Dietrich}, {Doroshenko},
  {Eastman}, {Efimov}, {Gaskell}, {Hedrick}, {Klimanov}, {Klimek}, {Kruse},
  {Lamb}, {Leighly}, {Minezaki}, {Nazarov}, {Petersen}, {Peterson},
  {Poindexter}, {Schlesinger}, {Sakata}, {Sergeev}, {Tobin}, {Unterborn},
  {Vestergaard}, {Watkins}, \& {Yoshii}}]{Denney09}
{Denney}, K.~D., {et~al.} 2009, \apj, 702, 1353

\bibitem[{{Dere} {et~al.}(2009){Dere}, {Landi}, {Young}, {Del Zanna},
  {Landini}, \& {Mason}}]{Dere09}
{Dere}, K.~P., {Landi}, E., {Young}, P.~R., {Del Zanna}, G., {Landini}, M., \&
  {Mason}, H.~E. 2009, \aap, 498, 915

\bibitem[{{Detmers} {et~al.}(2011){Detmers}, {Kaastra}, {Steenbrugge},
  {Ebrero}, {Kriss}, {Arav}, {Behar}, {Costantini}, {Branduardi-Raymont},
  {Mehdipour}, {Bianchi}, {Cappi}, {Petrucci}, {Ponti}, {Pinto}, {Ratti}, \&
  {Holczer}}]{Detmers11}
{Detmers}, R.~G., {et~al.} 2011, \aap, 534, A38

\bibitem[{{Fabian} {et~al.}(2009){Fabian}, {Zoghbi}, {Ross}, {Uttley}, {Gallo},
  {Brandt}, {Blustin}, {Boller}, {Caballero-Garcia}, {Larsson}, {Miller},
  {Miniutti}, {Ponti}, {Reis}, {Reynolds}, {Tanaka}, \& {Young}}]{Fabian09}
{Fabian}, A.~C., {et~al.} 2009, \nat, 459, 540

\bibitem[{{Fukumura} {et~al.}(2010){Fukumura}, {Kazanas}, {Contopoulos}, \&
  {Behar}}]{Fukumura10}
{Fukumura}, K., {Kazanas}, D., {Contopoulos}, I., \& {Behar}, E. 2010, \apj,
  715, 636

\bibitem[{{Halpern}(1984)}]{Halpern84}
{Halpern}, J.~P. 1984, \apj, 281, 90

\bibitem[{{Jones} {et~al.}(2011){Jones}, {McHardy}, {Moss}, {Seymour},
  {Breedt}, {Uttley}, {K{\"o}rding}, \& {Tudose}}]{Jones11}
{Jones}, S., {McHardy}, I., {Moss}, D., {Seymour}, N., {Breedt}, E., {Uttley},
  P., {K{\"o}rding}, E., \& {Tudose}, V. 2011, \mnras, 412, 2641

\bibitem[{{Kaastra} {et~al.}(2000){Kaastra}, {Mewe}, {Liedahl}, {Komossa}, \&
  {Brinkman}}]{Kaastra00}
{Kaastra}, J.~S., {Mewe}, R., {Liedahl}, D.~A., {Komossa}, S., \& {Brinkman},
  A.~C. 2000, \aap, 354, L83

\bibitem[{Kalberla {et~al.}(2005)Kalberla, Burton, Hartmann, Arnal, Bajaja,
  Morras, \& P{\"{o}}ppel}]{Kalberla05}
Kalberla, P. M.~W., Burton, W.~B., Hartmann, D., Arnal, E.~M., Bajaja, E.,
  Morras, R., \& P{\"{o}}ppel, W. G.~L. 2005, A\&A, 440, 775

\bibitem[{{Kallman} {et~al.}(2009){Kallman}, {Bautista}, {Goriely}, {Mendoza},
  {Miller}, {Palmeri}, {Quinet}, \& {Raymond}}]{Kallman09}
{Kallman}, T.~R., {Bautista}, M.~A., {Goriely}, S., {Mendoza}, C., {Miller},
  J.~M., {Palmeri}, P., {Quinet}, P., \& {Raymond}, J. 2009, \apj, 701, 865

\bibitem[{{Kallman} \& {McCray}(1982)}]{Kallman82}
{Kallman}, T.~R., \& {McCray}, R. 1982, \apjs, 50, 263

\bibitem[{{Kaspi} {et~al.}(2000){Kaspi}, {Brandt}, {Netzer}, {Sambruna},
  {Chartas}, {Garmire}, \& {Nousek}}]{Kaspi00}
{Kaspi}, S., {Brandt}, W.~N., {Netzer}, H., {Sambruna}, R., {Chartas}, G.,
  {Garmire}, G.~P., \& {Nousek}, J.~A. 2000, \apjl, 535, L17

\bibitem[{{Kaspi} {et~al.}(2002){Kaspi}, {Brandt}, {George}, {Netzer},
  {Crenshaw}, {Gabel}, {Hamann}, {Kaiser}, {Koratkar}, {Kraemer}, {Kriss},
  {Mathur}, {Mushotzky}, {Nandra}, {Peterson}, {Shields}, {Turner}, \&
  {Zheng}}]{Kaspi02}
{Kaspi}, S., {et~al.} 2002, \apj, 574, 643

\bibitem[{{King} {et~al.}(2011){King}, {Miller}, {Cackett}, {Fabian},
  {Markoff}, {Nowak}, {Rupen}, {G{\"u}ltekin}, \& {Reynolds}}]{King11}
{King}, A.~L., {et~al.} 2011, \apj, 729, 19

\bibitem[{{Kraemer} {et~al.}(2005){Kraemer}, {George}, {Crenshaw}, {Gabel},
  {Turner}, {Gull}, {Hutchings}, {Kriss}, {Mushotzky}, {Netzer}, {Peterson}, \&
  {Behar}}]{Kraemer05}
{Kraemer}, S.~B., {et~al.} 2005, \apj, 633, 693

\bibitem[{{Krongold} {et~al.}(2007){Krongold}, {Nicastro}, {Elvis},
  {Brickhouse}, {Binette}, {Mathur}, \& {Jim{\'e}nez-Bail{\'o}n}}]{Krongold07}
{Krongold}, Y., {Nicastro}, F., {Elvis}, M., {Brickhouse}, N., {Binette}, L.,
  {Mathur}, S., \& {Jim{\'e}nez-Bail{\'o}n}, E. 2007, \apj, 659, 1022

\bibitem[{{Kubota} {et~al.}(2007){Kubota}, {Dotani}, {Cottam}, {Kotani},
  {Done}, {Ueda}, {Fabian}, {Yasuda}, {Takahashi}, {Fukazawa}, {Yamaoka},
  {Makishima}, {Yamada}, {Kohmura}, \& {Angelini}}]{Kubota07}
{Kubota}, A., {et~al.} 2007, \pasj, 59, 185

\bibitem[{{Lee} {et~al.}(2001){Lee}, {Ogle}, {Canizares}, {Marshall}, {Schulz},
  {Morales}, {Fabian}, \& {Iwasawa}}]{Lee01}
{Lee}, J.~C., {Ogle}, P.~M., {Canizares}, C.~R., {Marshall}, H.~L., {Schulz},
  N.~S., {Morales}, R., {Fabian}, A.~C., \& {Iwasawa}, K. 2001, \apjl, 554, L13

\bibitem[{{Lobban} {et~al.}(2011){Lobban}, {Reeves}, {Miller}, {Turner},
  {Braito}, {Kraemer}, \& {Crenshaw}}]{Lobban11}
{Lobban}, A.~P., {Reeves}, J.~N., {Miller}, L., {Turner}, T.~J., {Braito}, V.,
  {Kraemer}, S.~B., \& {Crenshaw}, D.~M. 2011, \mnras, 414, 1965

\bibitem[{{Luketic} {et~al.}(2010){Luketic}, {Proga}, {Kallman}, {Raymond}, \&
  {Miller}}]{Luketic10}
{Luketic}, S., {Proga}, D., {Kallman}, T.~R., {Raymond}, J.~C., \& {Miller},
  J.~M. 2010, \apj, 719, 515

\bibitem[{{Magdziarz} {et~al.}(1998){Magdziarz}, {Blaes}, {Zdziarski},
  {Johnson}, \& {Smith}}]{Magdziarz98}
{Magdziarz}, P., {Blaes}, O.~M., {Zdziarski}, A.~A., {Johnson}, W.~N., \&
  {Smith}, D.~A. 1998, \mnras, 301, 179

\bibitem[{{Maitra} {et~al.}(2011){Maitra}, {Miller}, {Markoff}, \&
  {King}}]{Maitra11}
{Maitra}, D., {Miller}, J.~M., {Markoff}, S., \& {King}, A. 2011, \apj, 735,
  107

\bibitem[{{Mauche} {et~al.}(2001){Mauche}, {Liedahl}, \& {Fournier}}]{Mauche01}
{Mauche}, C.~W., {Liedahl}, D.~A., \& {Fournier}, K.~B. 2001, \apj, 560, 992

\bibitem[{{Mauche} {et~al.}(2003){Mauche}, {Liedahl}, \& {Fournier}}]{Mauche03}
---. 2003, \apjl, 588, L101

\bibitem[{{Mauche} \& {Raymond}(2000)}]{Mauche00}
{Mauche}, C.~W., \& {Raymond}, J.~C. 2000, \apj, 541, 924

\bibitem[{{Miller} {et~al.}(2006{\natexlab{a}}){Miller}, {Raymond}, {Fabian},
  {Steeghs}, {Homan}, {Reynolds}, {van der Klis}, \& {Wijnands}}]{Miller06}
{Miller}, J.~M., {Raymond}, J., {Fabian}, A., {Steeghs}, D., {Homan}, J.,
  {Reynolds}, C., {van der Klis}, M., \& {Wijnands}, R. 2006{\natexlab{a}},
  \nat, 441, 953

\bibitem[{{Miller} {et~al.}(2008){Miller}, {Raymond}, {Reynolds}, {Fabian},
  {Kallman}, \& {Homan}}]{Miller08}
{Miller}, J.~M., {Raymond}, J., {Reynolds}, C.~S., {Fabian}, A.~C., {Kallman},
  T.~R., \& {Homan}, J. 2008, \apj, 680, 1359

\bibitem[{{Miller} {et~al.}(2006{\natexlab{b}}){Miller}, {Raymond}, {Homan},
  {Fabian}, {Steeghs}, {Wijnands}, {Rupen}, {Charles}, {van der Klis}, \&
  {Lewin}}]{Miller06b}
{Miller}, J.~M., {et~al.} 2006{\natexlab{b}}, \apj, 646, 394

\bibitem[{{Nucita} {et~al.}(2010){Nucita}, {Guainazzi}, {Longinotti},
  {Santos-Lleo}, {Maruccia}, \& {Bianchi}}]{Nucita10}
{Nucita}, A.~A., {Guainazzi}, M., {Longinotti}, A.~L., {Santos-Lleo}, M.,
  {Maruccia}, Y., \& {Bianchi}, S. 2010, \aap, 515, A47+

\bibitem[{{Ogle} {et~al.}(2004){Ogle}, {Mason}, {Page}, {Salvi}, {Cordova},
  {McHardy}, \& {Priedhorsky}}]{Ogle04}
{Ogle}, P.~M., {Mason}, K.~O., {Page}, M.~J., {Salvi}, N.~J., {Cordova}, F.~A.,
  {McHardy}, I.~M., \& {Priedhorsky}, W.~C. 2004, \apj, 606, 151

\bibitem[{{Peterson} {et~al.}(2004){Peterson}, {Ferrarese}, {Gilbert}, {Kaspi},
  {Malkan}, {Maoz}, {Merritt}, {Netzer}, {Onken}, {Pogge}, {Vestergaard}, \&
  {Wandel}}]{Peterson04}
{Peterson}, B.~M., {et~al.} 2004, \apj, 613, 682

\bibitem[{{Ponti} {et~al.}(2006){Ponti}, {Miniutti}, {Cappi}, {Maraschi},
  {Fabian}, \& {Iwasawa}}]{Ponti06}
{Ponti}, G., {Miniutti}, G., {Cappi}, M., {Maraschi}, L., {Fabian}, A.~C., \&
  {Iwasawa}, K. 2006, \mnras, 368, 903

\bibitem[{{Porquet} \& {Dubau}(2000)}]{Porquet00}
{Porquet}, D., \& {Dubau}, J. 2000, \aaps, 143, 495

\bibitem[{{Pounds} \& {Vaughan}(2011)}]{Pounds11}
{Pounds}, K.~A., \& {Vaughan}, S. 2011, \mnras, 413, 1251

\bibitem[{{Proga} \& {Begelman}(2003)}]{Proga03}
{Proga}, D., \& {Begelman}, M.~C. 2003, \apj, 592, 767

\bibitem[{{Proga} {et~al.}(2000){Proga}, {Stone}, \& {Kallman}}]{Proga00}
{Proga}, D., {Stone}, J.~M., \& {Kallman}, T.~R. 2000, \apj, 543, 686

\bibitem[{Reynolds(1997)}]{Reynolds97}
Reynolds, C. 1997, MNRAS, 286, 513

\bibitem[{{Steenbrugge} {et~al.}(2009){Steenbrugge}, {Fenov{\v c}{\'{\i}}k},
  {Kaastra}, {Costantini}, \& {Verbunt}}]{Steenbrugge09}
{Steenbrugge}, K.~C., {Fenov{\v c}{\'{\i}}k}, M., {Kaastra}, J.~S.,
  {Costantini}, E., \& {Verbunt}, F. 2009, \aap, 496, 107

\bibitem[{{Young} {et~al.}(2005){Young}, {Lee}, {Fabian}, {Reynolds}, {Gibson},
  \& {Canizares}}]{Young05}
{Young}, A.~J., {Lee}, J.~C., {Fabian}, A.~C., {Reynolds}, C.~S., {Gibson},
  R.~R., \& {Canizares}, C.~R. 2005, \apj, 631, 733

\end{thebibliography}

\end{document}